\documentclass[twocolumn]{aastex63}
\usepackage[whole]{bxcjkjatype} % for arXiv (replace xeCJK)
\usepackage[utf8]{inputenc}
\accepted{to ApJ on September 16, 2022}
\shorttitle{Particle acceleration in BH magnetospheres}
\shortauthors{Hirotani et al.}
\graphicspath{{./}{figures/}}

\begin{document}

\title{Two-dimensional Particle-in-Cell simulations of axisymmetric black hole magnetospheres:
angular dependence of the Blandford-Znajek flux}

%% you desire. In v6.3 you can include a footnote in the title.

%% during compilation, LaTeX will do some limited checking of the format of
%% the ID to make sure it is valid. If the "orcid-ID.png" image file is 
%% present or in the LaTeX pathway, the OrcID icon will appear next to
%% the authors name.
%%
\correspondingauthor{Kouichi Hirotani, Hsien Shang}
\email{hirotani,shang@asiaa.sinica.edu}

\author[0000-0002-2472-9002]{Kouichi Hirotani}
\affiliation{Institute of Astronomy and Astrophysics, Academia Sinica, 
Taipei 10617, Taiwan, R.O.C.}

\author[0000-0001-8385-9838]{Hsien Shang （尚賢）}
\affiliation{Institute of Astronomy and Astrophysics, Academia Sinica, 
Taipei 10617, Taiwan, R.O.C.}

\author[0000-0001-5557-5387]{Ruben Krasnopolsky}
\affiliation{Institute of Astronomy and Astrophysics, Academia Sinica, 
Taipei 10617, Taiwan, R.O.C.}
% \collaboration{1}{(AAS Journals Data Scientists collaboration)}

\author[0000-0001-6031-7040]{Ken-ichi Nishikawa}
\affiliation{Department of Physics, Chemistry and Mathematics, 
Alabama A\&M University, Huntsville, AL 35811, USA
}
% \altaffiliation{}
% \affiliation{TeXnology Inc.}

% \collaboration{1}{(LaTeX collaboration)}

% \author{Michael Watson}
% \affiliation{Department of Physics, Fisk University, 
% 1000 17th Ave North, 
% Nashville, TN 37208, USA}
% % \affiliation{}

% \nocollaboration{2}



\begin{abstract}
We examine the temporary evolution of axisymmetric magnetospheres
around rapidly rotating black holes (BHs),
by applying our two-dimensional particle-in-cell simulation code.
Assuming a stellar-mass BH,
we find that the created pairs fail to screen the electric field
along the magnetic field,
provided that the mass accretion rate is much small compared to
the Eddington limit.
Magnetic islands are created by reconnection near the equator
and migrate toward the event horizon,
expelling magnetic flux tubes from the BH vicinity
during a large fraction of time.
When the magnetic islands stick to the horizon due to redshift
and virtually vanish,
a strong magnetic field penetrates the horizon,
enabling efficient extraction of energy from the BH.
During this flaring phase, a BH gap appears around the inner light surface
with a strong meridional return current toward the equator
within the ergosphere.
If the mass accretion rate is 0.025 percent of the Eddington limit,
the BH's spin-down luminosity becomes 16-19 times greater than 
its analytical estimate during the flares,
although its long-term average is only 6 percent of it.
We demonstrate that the extracted energy flux concentrates 
along the magnetic field lines 
threading the horizon in the middle latitudes.
It is implied that this meridional concentration of the Poynting flux
may result in the formation of limb-brightened jets from 
low-accreting BH systems.
\end{abstract}

\keywords{acceleration of particles --- 
magnetic fields --- methods: analytical --- methods: numerical ---
stars: black holes}


\section{Introduction}
\label{sec:intro}
The study of nonthermal plasmas in the vicinity of black holes (BHs) 
is astrophysically interesting
in the context of collimated relativistic outflows
observed from microquasars and active galactic nuclei.
Such relativistic outflows, a.k.a. jets,
are believed to be energized by rapidly rotating BHs
that are immersed in a globally ordered magnetic field.
In particular, when magnetic field ($\mbox{\boldmath$B$}$) lines 
thread the event horizon,
an electromotive force (EMF) is produced across the field lines
in the same way as an unipolar inductor.
In the direct vicinity of the horizon, 
this EMF induces a meridional current $\mbox{\boldmath$J$}$,
which exerts a counter torque on the BH 
via $\mbox{\boldmath$J$} \times \mbox{\boldmath$B$}$ force
\citep{bla77,Koide:2002:Sci,McKinney:2012:MNRAS}.
If such a current forms a closed circuit in a global BH magnetosphere, 
the extracted BH's rotational energy is carried away as the Poynting flux
to be dissipated at some electric load located at large distances, 
e.g., as synchrotron emissions in the jet downstream.

In the case of low-luminosity radio galaxies,
their flat spectrum radio emission with high brightness temperature
is interpreted to originate in a synchrotron-emitting jet
\citep{Jones:1974a:ApJ,Jones:1974b:ApJ,Blandford:1979:ApJ,Marscher:1983:ApJ,
       Lobanov:1998:A&A}.
In the case of BH binaries,
their flat spectrum radio emission during a hard/quiescent state 
is also considered to be from a jet
\citep{Hjellming:1988:ApJ,Stirling:2001:MNRAS,Dhawan:2000:ApJ,
       Fender:2004:MNRAS,Gallo:2005:MNRAS}.
In such low accreting systems, their small plasma density 
results in a negligible turbulent diffusion,
which prevents equatorial accretion to enter the jet-launching, 
polar regions in which horizon-penetrating magnetic field lines reside.

Although it is difficult to replenish jet materials by accretion
in this way, photon-photon pair production
is considered to be a viable mechanism as the source of jet plasmas.
Gap models envisage large-scale regions in which
a magnetic-field-aligned electric field accelerates
charged leptons into ultra-relativistic energies,
leading to a pair-production cascade of the gap-emitted $\gamma$-rays
in a target, soft photon field
\citep{bes92, Hirotani:1998:ApJ, Levinson:2000:PhRvL, levi11, Ptitsyna:2016:A&A, 
       Hirotani:2016:ApJ, Hirotani:2017:ApJ, Levinson:2017:PhRvD, 
       Hirotani:2018:ApJ, Ford:2018:PhRvD,
       Katsoulakos:2020:ApJ}.
Drizzle models, on the other hand,
consider near-horizon, transient and small scale regions
in which high-energy component of electrons produce MeV photons,
which collide each other to produce pairs
\citep{Moscibrodzka:2011:ApJ,Wong:2021:ApJ}.
In the present paper, we focus on the former model,
considering a jet-launching, low-accreting BH systems,
which are realized in the center of low-luminosity radio galaxies
or BH binaries in a hard/quiescent state.

When a BH is accreting at a highly sub-Eddington rate,
such as in the case of M87*
or BH binaries in a hard/quiescent state,
plasmas become highly collisionless,
which makes it impossible to justify the magnetohydrodynamic (MHD) 
approximations in the jet-launching regions
\citep[][hereafter Paper~I]{Hirotani:2021:ApJ}.
Instead, kinetic method, or the particle-in-cell (PIC) 
method becomes appropriate
(e.g., \cite{Nishikawa:2021:LRCA} for a recent review).

Assuming that the gap local physics does not affect the global structure, 
such as the magnetospheric currents,
one-dimensional general relativistic PIC (GRPIC) simulations were performed by
\citet{Levinson:2018:AA,Chen:2020:ApJ,Kisaka:2020:ApJ}.
They consistently solved the radiative-transfer equation
(or at least treated the inverse-Compton scatterings
 and pair production in a realistic way in the last case),
together with the motion of the created pairs and the evolution of 
the electromagnetic fields.

To incorporate the back reaction on the global magnetosphere, however,
we should proceed to two-dimensional (2D) cases.
In this context, \citet{Parfrey:2019:PhRvL} first performed
2D GRPIC simulations of BH magnetospheres,
assuming the pair injection rate is proportional to
the strength of acceleration electric field,
instead of solving the radiative transfer equation.
Adopting an extremely small magnetic field strength,
they demonstrated that the Penrose process contributes
to the extraction of energy from a maximally rotating BH.
Subsequently, \citet{Crinquand:2020:PhRvL} examined 2D GRPIC simulations,
solving the radiative transfer equation,
and assuming a fixed monopole magnetic field, 
which is, indeed, appropriate in a time-averaged sense
in the horizon vicinity.
They showed that the BH's rotational energy can be electromagnetically 
extracted via the Blandford-Znajek (BZ) process, and that
a highly time-dependent spark gap opens near the inner light surface.
In addition, \citet{Crinquand:2021:A&A} coupled their 2D GRPIC code
with a raytracing algorithm by post processing, and
examined synthetic $\gamma$-ray light curves of the gap activity.
Moreover, \citet{Bransgrove:2021:PhRvL} applied 2D GRPIC
and 3D MHD methods to a stellar-mass BH which is collapsing from 
a neutron star surrounded by plasma,
and demonstrated that the \lq no hair' theorem holds
in the sense that the stress-energy tensor decays exponentially in time.

In addition to these works, 
in Paper~I, we applied our 2D GRPIC code % paper
to stellar-mass BHs, without solving the radiative transfer equation.
When solving the Maxwell equations, we solved only three components of 
the electromagnetic fields, namely the radial and meridional components of
the electric field and the toroidal component of the magnetic field.
Assuming a radial magnetic-field geometry near the BH,
we demonstrated that the BH's rotational energy is preferentially extracted
along the magnetic field lines threading the event horizon
in the middle latitudes, namely between $60^\circ$ and $70^\circ$
(or $110^\circ$ and $120^\circ$) from the rotation axis.

Developing the 2D GRPIC method adopted in Paper~I,
we solve all the six components of the electromagnetic fields
in the present paper.
In the next section, we describe the basic equations in our GRPIC scheme.
Then in \S~\ref{sec:nonstationary},
we show that the main conclusion of Paper~I
-- the middle-latitude concentration of the BZ flux --
also holds when we solve all the six electromagnetic-field components,
and demonstrate that the force-free approximation breaks down
when plasmas are less efficiently supplied in the magnetosphere.
We finally discuss an implication on the formation of a limb-brightened jet
in \S~\ref{sec:disc}.

\section{The particle-in-cell (PIC) scheme}
\label{sec:PIC}
We formulate our axisymmetric, 2D GR PIC method in this section. 

\subsection{Background geometry} 
\label{sec:background}
Around a rotating, non-charged BH, 
the background geometry is described by the Kerr metric
\citep{kerr63}.
In the Boyer-Lindquist coordinates \citep{boyer67}, 
the line element can be expressed as
\begin{equation}
 ds^2= g_{tt} dt^2
      +2g_{t\varphi} dt d\varphi
      +g_{\varphi\varphi} d\varphi^2
      +g_{rr} dr^2
      +g_{\theta\theta} d\theta^2,
  \label{eq:metric}
%  \eqno{(S1)},
\end{equation}
where 
\begin{equation}
   g_{tt} 
   \equiv 
   -\frac{\Delta-a^2\sin^2\theta}{\Sigma},
   \qquad
   g_{t\varphi}
   \equiv 
   -\frac{2Mar \sin^2\theta}{\Sigma}, 
  \label{eq:metric_2}
%  \eqno{(S2)}
\end{equation}
\begin{equation}
   g_{\varphi\varphi}
     \equiv 
     \frac{A \sin^2\theta}{\Sigma} , 
     \qquad
   g_{rr}
     \equiv 
     \frac{\Sigma}{\Delta} , 
     \qquad
   g_{\theta\theta}
     \equiv 
     \Sigma ;
  \label{eq:metric_3}
%  \eqno{(S1)}
\end{equation}
$\Delta \equiv r^2-2Mr+a^2$,
$\Sigma\equiv r^2 +a^2\cos^2\theta$,
$A \equiv (r^2+a^2)^2-\Delta a^2\sin^2\theta$.
In equations~(\ref{eq:metric})--(\ref{eq:metric_3}), 
we adopt the geometrized unit, putting $c=G=1$, where
$c$ and $G$ denote the speed of light and the gravitational constant,
The horizon radius, $r_{\rm H} \equiv M+\sqrt{M^2-a^2}$,
is obtained by $\Delta=0$, 
where $r_{\rm g}= GM c^{-2}= M$ 
corresponds to the gravitational radius.
The spin parameter becomes $a=M$ for a maximally rotating BH,
and $a=0$ for a non-rotating BH.

To avoid singular behaviours 
at $\sin\theta=0$ (i.e., at the poles)
due to the differential operator
$\csc\theta \partial_\theta$ in the Maxwell equations,
we introduce a new meridional variable, $y \equiv 1-\cos\theta$.
Adopting this $y$ coordinate, we obtain
\begin{equation}
  \frac{1}{\sin\theta}\frac{\partial}{\partial\theta}
  = \frac{\partial}{\partial y}.
  \label{eq:xi}
\end{equation}
Here, $y=0$ (or $y=2$) corresponds to the rotational axis 
in the uuper (or lower) hemisphere,
and $y=1$ denotes the equatorial plane.

In the radial direction, we adopt the so-called
\lq\lq tortoise coordinate'' $r_\ast$,
\begin{equation}
  \frac{dr_\ast}{dr} 
  \equiv \frac{r^2+a^2}{\Delta}.
  \label{eq:def_tortoise}
\end{equation}
In this coordinate, the event horizon corresponds to 
$r_\ast \rightarrow -\infty$.
Away from the BH ($r \gg M$),
it tends to the standard radial coordinate,
$dr_\ast / dr \rightarrow 1$.

%
% Let us look briefly at the collisionless nature of the plasmas
% in \S~\ref{sec:collisionless},
% before turning to a closer examination of the temporal evolution
% of the BH magnetosphere in the rest of this section.

\subsection{Background electromagnetic fields} 
\label{sec:background_EM}
Throughout this paper, 
we assume that there exist stationary electromagnetic fields
that are represented by the Wald solution
\citep{Wald:1974:PhRvD},
which is realized when a ring current flows on the equator 
at a large enough distance from the BH.
The amplitude of the ring current determines
the magnetic field strength, $B$, near the BH.
To specify $B$, we use its equipartition value, $B_{\rm eq}$,
with the accreting plasmas.
When accretion takes place as 
an advection-dominated accretion flow (ADAF)
\citep{ichimaru77,narayan94,Tchekhovskoy:2011:MNRAS,Narayan:2021:arXiv},
we obtain 
\citep{Yuan:2014:ARA&A},
\begin{equation}
  B_{\rm eq}(r)
  = 9.7 \times 10^7 
    \left( \frac{\dot{m}}{M_1} \right)^{1/2} 
    \left( \frac{r}{2M} \right)^{-5/4} 
    \mbox{ G}.
  \label{eq:B_eq}
\end{equation}
Here, the dimensionless accretion rate 
$\dot{m}$ is defined by
\begin{equation}
  \dot{m} \equiv \frac{\dot{M}}{\dot{M}_{\rm Edd}},
\end{equation}
where $\dot{M}$ denotes the mass accretion rate.
The Eddington accretion rate is defined by
\begin{equation}
  \dot{M}_{\rm Edd} 
  \equiv 
  \frac{L_{\rm Edd}}{\eta_{\rm eff} c^2}
  = 1.39 \times 10^{19} M_1 {\rm g \ s}^{-1},
\end{equation}
where $L_{\rm Edd}$ denotes the Eddington luminosity.
We adopt the conversion efficiency $\eta_{\rm eff}=0.1$.

In the present paper, 
we assume that the background, Wald solution has 
a field strength $B$
so that its meridional average at $r=2M$ may match 
$B_{\rm eq}$.
Since our code cannot simulate a plasma containing protons,
we do no solve the normal plasmas consisting an ADAF.
Instead, as the footpoint of a pair-dominated jet
\citep{Reynolds:1996:MNRAS,Wardle:1998:Natur,
Hirotani:2005:ApJ,Kino:2014:ApJ},
we focus on the electron-positron pair plasmas
that are created within the magnetosphere.
Therefore, we use $\dot{m}$ merely to
specify $B_{\rm eq}$ (and hence $B$ near the BH),
and to specify the pair-supply rate, $\dot{N}_\pm$ (\S~\ref{sec:supply}).

\subsection[]{The Maxwell equations}
\label{sec:Maxwell}
In addition to the Wald solution described in the foregoing section, 
we consider additional electromagnetic fields
produced by the electric currents flowing near the BH.
Accordingly, the total electromagnetic fields are given by
the superposition of the stationary Wald solution
and these non-stationary fields.
We describe how to solve the latter fields 
within our PIC scheme below.

When we solve the Maxwell equations, 
we should not solve for the Faraday tensor components
that are defined in the Boyer-Lindquist coordinates,
because numerical instabilities arise inside the ergosphere,
in which a non-rotating observer becomes space-like.
Instead, we can adopt a physical observer,
such as the Zero-Angular-Momentum Observer (ZAMO)
to avoid this ill behaviour.
In this paper, 
instead of converting quantities into their ZAMO-measured values, 
we redefine the $r$ and $\theta$ 
components of the electric field such that
\begin{eqnarray}
  E_1 
  &\equiv& F_{r t}-\omega F_{\varphi r}, 
  \label{eq:Ep1} \\
  E_2 
  &\equiv& (F_{\theta t}+\omega F_{\theta \varphi}) / r_{\rm g}, 
  \label{eq:Ep2}    
\end{eqnarray}
where $F_{\mu \nu}$ denotes a Faraday tensor component, and
$\omega= 2Mar/A$ does the frame-dragging angular frequency.
These transformations mimic the adoption of ZAMO;
see also the arguments after equation~(12) of Paper~I 
for physical explanations of these variable transformations.
In what follows, 
we put $r_{\rm g}=1$ in equations for simplicity,
and recover $r_{\rm g}$ when appropriate.

Using these re-defined electric field 
(eqs.~[\ref{eq:Ep1}] \& [\ref{eq:Ep2}]),
we find the following six time-dependent Maxwell equations:
\begin{equation}
  \frac{\partial B_1}{\partial t}
  = -\frac{\partial E_3}{\partial y},
  \label{eq:Maxwell_1}
\end{equation}
\begin{equation}
  \frac{\partial B_2}{\partial t}
  = \frac{\partial E_3}{\partial r_\ast},
  \label{eq:Maxwell_2}
\end{equation}
\begin{eqnarray}
  \frac{\partial B_3}{\partial t}
  &=& \frac{\Delta\sin\theta}{\Sigma} 
       \frac{\partial E_1}{\partial y}
      -\frac{r^2+a^2}{\Sigma}
       \frac{\partial E_2}{\partial r_\ast}
  \nonumber \\
  &+& \frac{r^2+a^2}{\Sigma} 
       \left(     \frac{\Delta \sin\theta}{r^2+a^2}
              B_1 \frac{\partial \omega}{\partial r}
             +B_2 \frac{\partial \omega}{\partial \theta}
       \right),
  \label{eq:Maxwell_3}
\end{eqnarray}
\begin{equation}
  \frac{\partial E_1}{\partial t}
  = \frac{A}{\Sigma} \frac{\partial (B_3 \sin\theta)}
                          {\partial y}
    -4 \pi \frac{\Sigma^2}{A} J_1
  \label{eq:Maxwell_4}
\end{equation}
\begin{equation}
  \frac{\partial E_2}{\partial t}
  = -\frac{(r^2+a^2) \Sigma}{A} 
     \frac{\partial B_3}{\partial r_\ast}
    -4 \pi \frac{\Delta \Sigma^2}{A} J_2
  \label{eq:Maxwell_5}
\end{equation}
\begin{eqnarray}
  \frac{\partial E_3}{\partial t}
  = && \frac{2 M a (r^2+a^2) \sin^2\theta}{\Sigma}
       \frac{\partial}{\partial r_\ast} 
       \left[ \frac{r}{\Sigma} E_1 \right]
    \nonumber\\
    &+& \frac{r^2+a^2}{\Sigma}
        \frac{\partial}{\partial r_\ast}
        \left[ \frac{(r^2+a^2)\Sigma}{A} B_2 \right]
    \nonumber\\
    &+& \frac{2Mar\sin^2\theta}{\Sigma}
        \frac{\partial}{\partial y}
        \left[ \frac{\sin\theta}{\Sigma} E_2 \right]
    \nonumber\\
    &-& \frac{\sin^2\theta}{\Sigma}
        \frac{\partial}{\partial y}
        \left[ \frac{\Delta\Sigma}{A} B_1 \right]
    \nonumber\\
    &-& 4\pi \Delta \sin^2\theta \cdot J_3,
    \label{eq:Maxwell_6}
\end{eqnarray}
where
\begin{equation}
  B_1 
  \equiv \frac{F_{\theta \varphi}}{r_{\rm g}^2 \sin\theta},
  \label{eq:Br}
\end{equation}
\begin{equation}
  B_2 
  \equiv \frac{\Delta}{r^2+a^2} \frac{F_{\varphi r}}{r_{\rm g}},
  \label{eq:Bth}
\end{equation}
\begin{equation}
  B_3 
  \equiv \frac{F^r{}_{\theta}}{r_{\rm g}},
  \label{eq:Bph}
\end{equation}
and
\begin{equation}
  E_3 \equiv \frac{F_{\varphi t}}{r_{\rm g}},
  \label{eq:Bph}
\end{equation}
if we recover $r_{\rm g}$ for clarity.
The electric currents are defined by
\begin{equation}
  J_1 \equiv \frac{r_{\rm g}}{c} J^r
      = \frac{r_{\rm g}}{c} 
        \sum_{n=1}^{m} \frac{q_n}{\delta V} c
        \frac{u^r}{u^t},
  \label{eq:def_J1}
\end{equation}
\begin{equation}
  J_2 \equiv \frac{r_{\rm g}^2}{c} J^\theta
      = \frac{r_{\rm g}}{c} 
        \sum_{n=1}^{m} \frac{q_n}{\delta V} c
        \frac{u^\theta}{u^t},
  \label{eq:def_J1}
\end{equation}
\begin{equation}
  J_3 \equiv \frac{r_{\rm g}^2}{c} J^\varphi
      = \frac{r_{\rm g}}{c} 
        \sum_{n=1}^{m} \frac{q_n}{\delta V} c
        \frac{u^\varphi}{u^t},
  \label{eq:def_J1}
\end{equation}
where the sigma symbol, $\sum_{n=1}^{m}$,
denotes a summation of the electric currents
carried by all the particle crossing the area
where we count the current,
$m$ designates the number of macro particles in each subdomain,
$q_n$ does the charge on the macro particle
with identity number $n$,
and $\delta V$ does the invariant three-dimensional volume
of each cell around the grid point.
The four-velocity components, 
$u^t$, $u^r$, $u^\theta$, and $u^\varphi$
satisfy the definition of the proper time, 
$g_{\mu\nu} u^\mu u^\nu = -1$.
Note that $u^r/u^t= (1/c) dr/dt$, 
$u^\theta/u^t= (r_{\rm g}/c) d\theta/dt$, and 
$u^\varphi/u^t= (r_{\rm g}/c) d\varphi/dt$
are all dimensionless.

It is worth noting that solving 
equations~(\ref{eq:Maxwell_1})-(\ref{eq:Maxwell_6})
for the PIC fields, $E_i$ and $B_i$ ($i=1,2,3$),
is equivalent with solving the Maxwell equations
for the total electromagnetic fields,
because the Maxwell equations are linear, and hence additive.
For instance, writing the total fields
as a summation of the (stationary) Wald fields 
and the (time-dependent) PIC fields,
we find that all the terms containing the Wald fields vanish.
Namely, temporal derivatives of the Wald fields vanish
by definition.
The curl of the Wald electric field vanishes by the Faraday's law.
The curl of the Wald magnetic field vanishes by the Ampere's law,
because we impose a current-free condition
when obtaining the Wald solution.
Note that the required equatorial, ring current at a large enough radius
comes into the Wald solution only through the outer boundary condition,
not through the (electric-current) source terms.

Let us also describe how to construct the electric currents.
We divide the particle motion into the poloidal and toroidal components,
adopting the area weighting \citep{villa92} on the poloidal plane,
and computing the toroidal component of the current from the azimuthal
displacement of individual charged leptons, reflecting the area weighting.
However, the area weighting in a curved spacetime results in
an accumulation of small numerical errors in each time step,
and eventually ends up with non-physical short-wavelength noise
in the electromagnetic fields
due to failure to satisfy the Gauss's law.
Therefore, in the present paper,
we apply the Poisson correction to the electric field
\citep[e.g.,][]{Langdon:1976:cofu}.

Let us briefly describe the units used in the code.
Electromagnetic field components
$E_1$, $E_2$, $E_3$, $B_1$, $B_2$, $B_3$, 
are computed in the cgs gaussian unit.
Note that all these six components are well-behaved
at the horizon.
The charge density $\rho_{\rm e}$ is also measured 
in the cgs gaussian unit (i.e., $\mbox{statcoulomb cm}^{-3}$).
The current components, $J_1$, $J_2$, and $J_3$ are
in $\mbox{statampere} \times (r_{\rm g}/c)$ unit.
The macro particle's four-velocity components are dimensionless.
Namely, $u^0= dt / d\tau$ denotes the ratio of the elapsed 
coordinate time $dt$ and the proper time $d\tau$.
Thus, $u^0$ became the Lorentz factor 
in the special relativistic limit.
$u^1= dr / d\tau$ denotes the radial velocity in $r$ (not $r_\ast$)
coordinate in $d\tau$ basis,
$u^2= d\theta / d\tau$ does the meridional angular velocity, and
$u^3= d\varphi / d\tau$ does the azimuthal angular velocity.

To solve these six Maxwell equations
(eqs.~[\ref{eq:Maxwell_1}]--[\ref{eq:Maxwell_6}]),
we must impose boundary conditions.
Along the northern and southern polar axes 
(i.e., at $\theta=0$ and $\theta=\pi$),
we impose
\begin{eqnarray}
  && B_2 = 0, \quad 
     B_3 = 0, \quad
     F_{\theta t} = 0, \quad
     E_3= 0,
  \nonumber\\
  &&
  \frac{\partial B_1}{\partial y} = 0, 
  \frac{\partial F_{r t}}{\partial y}= 0.
  \label{eq:BDC_1}
\end{eqnarray}
% Let us briefly consider the meaning of $\partial_y B_1=0$.
% The radial component of the magnetic field is given by
% \begin{equation}
%   B^r
%   = \frac{1}{\Sigma}
%     \left( \frac{\Delta-a^2 \sin^2\theta}{\Sigma} 
%            \frac{F_{\theta\varphi}}{\sin\theta}
%           -\frac{2Mar}{\Sigma}\sin\theta F_{\theta t}
%     \right).
%   \label{eq:def_Br}
% \end{equation}
% Note that $B^r$ should be an even function of $\theta$
% around the north pole by symmetry.
% Thus, in the leading order in $\theta$ expansion around $\theta = 0$,
% $B_1= F_{\theta\varphi}/\sin\theta$
% should be an even function,
% because $\Sigma$, $\sin^2\theta$ and $\sin\theta F_{\theta t}$ 
% are even functions there.
% By the same argument, $B_1= F_{\theta\varphi}/\sin\theta$
% should be an even function also at the south pole.
% Therefore, we can impose $\partial_y B_1=0$ at both poles.

At the outer boundary,
we impose that the radial derivatives of 
$E_1$, $E_2$, $E_3$, $B_1$, $B_2$, and $B_3$
vanish for simplicity.
We tested several combinations of other radial boundary conditions,
and confirmed that conclusions were unaffected.

At the inner boudary,
we impose the \lq\lq radiative boundary condition''
so that the tangential components of the electromagnetic fields
may look to a ZAMO like radiation propagating into the stretched horizon,
or equivalently,
all the components of the electromagnetic fields should be finite
for a freely-falling observer
\cite[\S~III~C~4 of][]{thorne86}.
In ZAMO's orthonormal basis, this condition reads
$ (0,B^{\hat \theta},B^{\hat \varphi})
 =(0,E^{\hat \theta},E^{\hat \varphi)} \times (1,0,0) $,
which gives the following inner boundary conditions
\begin{eqnarray}
  E_3 &=& \frac{r^2+a^2}{\sqrt{A}} B_2
  \nonumber\\
  B_3 &=& -\frac{\sqrt{A}}{\Sigma} E_2 
  \label{eq:innerBDC}
\end{eqnarray}
in our current notation.

\subsection[]{Initial conditions}
\label{sec:initial}
We start the PIC simulation from
\begin{equation}
  E_1= E_2= E_3= B_1= B_2= B_3=0.
  \label{eq:initial_EM}
\end{equation}
at time $t=0$. 
Note that $E_i$ and $B_i$ ($i=1,2,3$)
describe the nonstationary fields that are to be
superposed on the stationary Wald fields.
We assume that the densities of electrons and positrons
are $0.1 n_{\rm GJ}$ at $t=0$,
where the Goldreich-Julian number density
is defined by 
\begin{equation}
  n_{\rm GJ}(r,\theta)
  = \frac{\omega_{\rm H} B(r,\theta)}
         {4 \pi c e}
  \label{eq:nGJ}
\end{equation}
at each point.
In each cell, we give 50 initial macro electrons and positrons
(i.e., 100 in total) with equal charge weight,
distributing their positions randomly in the cell.
We assume that these macro particles are static 
on the poloidal plane and corotating with the space time at $t=0$. 
That is, macro electrons and positrons are moving at the same
toroidal velocity initially;
accordingly, they do not carry any electric currents at $t=0$.

\subsection[]{Particle equation of motion}
\label{sec:EOM}
In a highly vacuum BH magnetosphere,
charged leptons are deccelerated by 
the radiation-reaction forces.
In the same manner as in Paper~I,
we include the radiation reaction force
as a friction term in the equation of motion (EOM).

With such a friction term, the EOM can be expressed by
\citep[Chapter~17 of ][]{jackson62}
\begin{eqnarray}
  \frac{du^\mu}{d\tau}
  =
  &&
  -\Gamma^\mu{}_{\nu\rho} u^\nu u^\rho
    + \frac{e}{m} F^\mu{}_\nu u^\nu
  \nonumber\\
  && 
  +\frac23 \frac{e^2}{m}
  \left( \frac{d^2 u^\mu}{d\tau^2}
        -u^\mu \frac{du^\nu}{d\tau} \frac{du_\nu}{d\tau} \right)
  \label{eq:EOM}
\end{eqnarray}
where $d\tau$ refers to the particle proper time, 
$e$ represents the absolute value of 
the charge on the electron,
and $m$ does the mass of the electron.

Let us briefly describe the boundary conditions on
the motion of electrons and positrons.
Due to the symmetry, 
we assume that the particles moving across the polar axis
(at $\theta=0$ or $\pi$) will be reflected 
toward the equator with opposite meridional velocity.
Both the inner and outer boundaries are treated as particle sinks.
Thus, when particles move across these two radial boundaries,
they are excluded from the simulation.

\subsection[]{Plasma supply}
\label{sec:supply}
In BH magnetospheres, pairs can be supplied via 
two-photon and/or one-photon (i.e., magnetic) 
pair production processes.
In the present paper, we focus on 
the former process, and consider the collisions
of MeV photons emitted via Bremsstrahlung from an ADAF.
% In subsequent papers, we will also consider
% the collisions between the gap-emitted (inverse-Compton)
% photons and the ADAF-emitted (synchrotron) photons.
The collision rate,
or equivalently, the pair supply rate (pairs per second per volume)
is given by
\begin{equation}
    \dot{N}_\pm = c \sigma_{\gamma\gamma} n_\gamma{}^2,
    \label{eq:dotN}
\end{equation}
where $\sigma_{\gamma\gamma}$ denotes the total cross section 
of photon-photon pair production, and 
$n_\gamma$ does the MeV photon density.
Adopting the Newtonian self-similar ADAF model
\citep{Mahadevan:1997:ApJ},
and assuming that the most energetic MeV photons
are emitted within $r=4M$,
we obtain
(Paper~I)
\begin{equation}
  \dot{N}_\pm
  \approx 1.0 \times 10^{24}
    \dot{m}^4 M_1{}^{-2}
    \max \left[ \left(\frac{r}{4M}\right)^{-4}, 1 \right]
  \label{eq:dotN2}
\end{equation}

We randomly introduce a macro particle in each cell 
at every time step with probability 
$1/k_{\rm create}=0.01$;
that is, particles are injected in each cell 
at every $k_{\rm create}=100$ time steps on average.
In this case, each created macro positron or electron
has the electric charge
\begin{equation}
    q_i= \pm e \dot{N}_\pm k_{\rm create} \Delta_t \Delta_V,
\end{equation}
where $\Delta_t$ denotes the interval of each time step, and
$\Delta_V$ the invariant volume of each cell.
Note that 
$\Delta_t \Delta_V= \sqrt{-g} dt dr d\theta d\varphi
 = 2\pi \sqrt{-g} \Delta_t \Delta_r \Delta_\theta$
holds, 
where $\Delta_r$ and $\Delta_\theta$ denote
the intervals in Boyer-Linquist 
radial and meridional coordinates.
In the present paper, $r_\ast$ and $y$ coordinates are
uniformly gridded; thus, 
both $r$ and $\theta$ are gridded non-uniformly.

As the PIC simulation proceeds, 
the number of macro particles
increases with $t$ to saturate at a few hundred 
in each PIC cell on average.
Here, the maximum value of the Courant number is set
to be $0.5$ in $r_\ast$ and $y=1-\cos\theta$ coordinates.
In total, there are about $3 \times 10^8$ 
macro particles in the entire simulation region.

To solve the temporal evolution of the electromagnetic fields
and the particle distribution functions,
we adopt a radial grid of 1200 uniform cells 
(in $r_\ast$ coordinate) between
$1.01 r_{\rm H} < r < r_{\rm out}= 25.0M$,
and 1120 uniform cells 
(in $y$ coordinate) between $0 < y=1-\cos\theta < 2$,
which corresponds to $0^\circ < \theta < 180^\circ$.
It is noteworthy that the plasma skin depth can {\it not} be resolved 
at $r>7M$ when the plasma density increases enough 
(particularly in the polar regions).
Thus, we adopt only the results obtained in $r < 7M$ 
as the appropriate PIC solution,
interpreting that the solution in $r > 7M$ merely gives
the outer boundary condition at $r = 7M$.
Alternatively, we could set some outer boundary conditions 
on the electromagnetic fields and the particle injection rate
at $r=7M$.
Nevertheless, in the present paper,
we infer the boundary conditions at $r=7M$
by solving the magnetosphere also in $7M < r < r_{\rm out}$.
By this treatment, the boundary conditions on the electromagnetic fields
at $r=r_{\rm out}$ little affect the solution in $r<7M$.
On the other hand, inward particle flux at $r=7M$ 
do depend on the condition imposed on the particle injection rate 
at $r=r_{\rm out}$.
Thus, we assume no particle injection across $r=r_{\rm out}$,
although particles freely escape outward across $r=r_{\rm out}$.
Since the pair production rate rapidly decreases with radius 
(eq.~[\ref{eq:dotN2}]),
particle inward flux at $r=7M$ is little affected
by the outer boundary position,
as long as it is located at $r \ge 25M$.

We adopt the cgs Gaussian unit in the code.
Thus, in the Maxwell equations (\ref{eq:Maxwell_1})--(\ref{eq:Maxwell_6}),
we measure the electric field $E_i$ ($i=1,2,3$) 
in $\mbox{statvolts cm}^{-1}$,
and the magentic field $B_i$ in gauss.
Time and spatial variables are in dimensionless unit.
For instance, $t$ is measured in $r_{\rm g} c^{-1}$, and
$r$ is in $r_{\rm g}$;
$y \equiv 1-\cos\theta$ is dimensionless by definition.
In the particle equations of motion (\ref{eq:EOM}),
the four velocity $u^\mu$ is measured in $c$,
and the proper time $\tau$ in $r_{\rm g} c^{-1}$.

It is checked {\it a posteriori} 
that the invariant grid intervals resolve 
the skin depth
\begin{equation}
 l_{\rm p}= \frac{c}{\omega_{\rm p}},
 \label{eq:skin_depth}
\end{equation}
at every point at any elapsed time,
where the plasma frequency $\omega_{\rm p}$ is computed by the plasma density
$n_\pm$ and its mean Lorentz factor $\langle\gamma\rangle$ as
\citep{Levinson:2018:AA}
\begin{equation}
  \omega_{\rm p}= \sqrt{\frac{4\pi e^2 n_\pm}{m \langle \gamma \rangle}}.
  \label{eq:plasma_freq}
\end{equation}
In addition, we adopt a heavy electron mass of
$m= m_{\rm p}/20$.
Partly because we adopt this heavy electron/positron mass,
and mainly because the leptons are ultra-relativistic  
($\langle\gamma\rangle \gg 1$),
and because the plasma density is very small ($n_\pm \sim n_{\rm GJ}$),
the plasma skin depth, $l_{\rm p}$, 
is resolved by the current grid intervals
for stellar-mass BHs
(\S~\ref{sec:skin}).

\section[]{Nonstationary magnetosphere}
\label{sec:nonstationary}
In this section, we apply the PIC method to
a rapidly rotating stellar-mass BH
with spin $a=0.9M$ and mass $M=10 M_\odot$.
We adopt a small mass accretion rate
$\dot{m}=2.5 \times 10^{-4}$ throughout this paper.
% After describing the electric field
% in \S\S~\ref{sec:E_field_init}--\ref{sec:E_field},
% we examine the magnetic field in \S~\ref{sec:B_field}.
% Then we focus on the angular dependence of the BZ flux 
% in \S~\ref{sec:BZ_flux},
% and its power spectra in \S~\ref{sec:BZflux_PSD}.
In what follows, electromagnetic fields mean
the total fields that are
obtained by superposing the stationary Wald fields
and the nonstationary PIC fields,
unless explicitly mentioned to distinguish them.

\subsection{Initial electromagnetic fields}
\label{sec:E_field_init}
\begin{figure*}
%\vspace*{-16.0truecm}
\includegraphics[width=\textwidth]{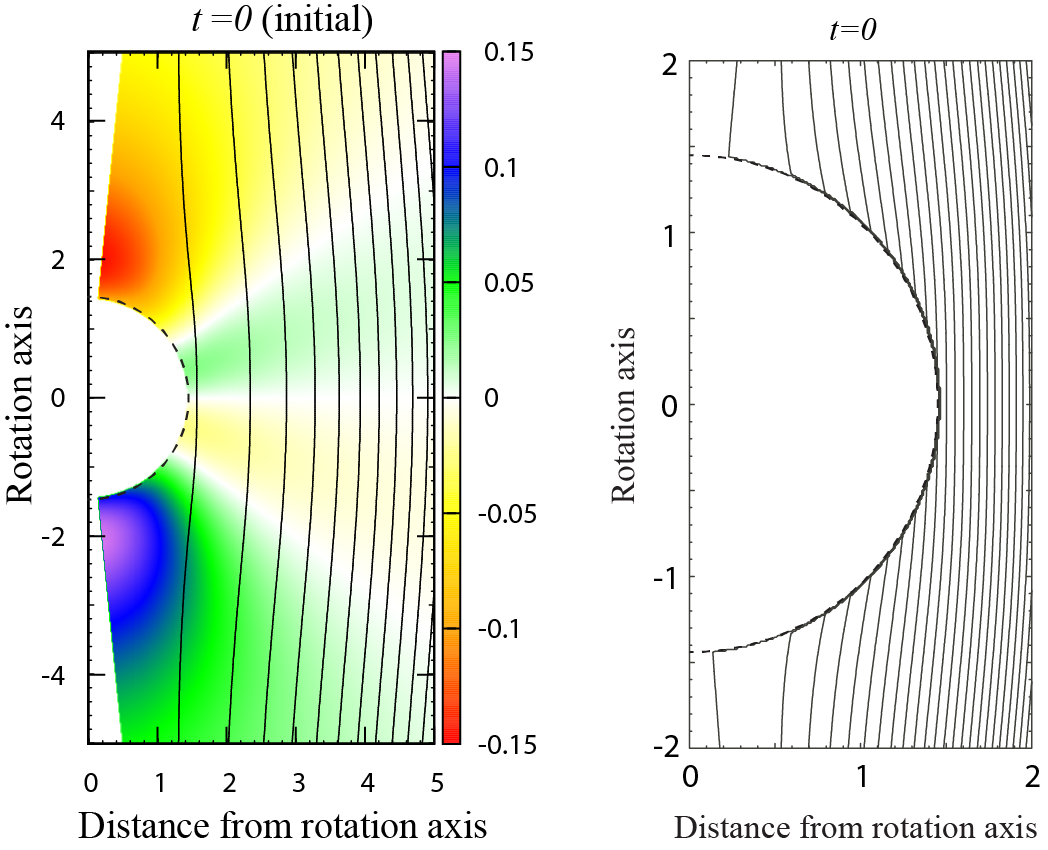}
\caption{
{\it Left:}
Initial distribution of 
$\mbox{\boldmath$E$} \cdot \mbox{\boldmath$B$} 
 / B_{\rm eq}(2M){}^2$ 
(color image)
on the poloidal plane ($r$,$\theta$),
when the BH's mass and spin parameter is 
$M=10M_\odot$ and $a=0.9M$, 
and the dimensionless mass accretion rate is
$\dot{m}=2.5 \times 10^{-4}$.
Black solid curves denote equi-$A_\varphi$ contours,
which give magnetic field lines in a stationary and axisymmetric
magnetosphere.
The magnetic-field strength 
is chosen to be $B(r=2M)=B_{\rm eq}(r=2M)$
after averaging over the spherical surface at $r=2M$
(see text).
The electromagnetic fields are given by the Wald solution,
which is stationary and axisymmetric.
Both the abscissa and ordinate are measured in $GMc^{-2}$ unit.
The event horizon is located at $r=1.435M$,
and represented by the dashed semicircle.
{\it Right:}
Close up of the magnetic field lines in the BH vicinity.
}
    \label{fig:Epara_1}
\end{figure*}

Because of the frame dragging, there appears an electric field
along the local magnetic field lines.
For $\dot{m}= 0.00025$,
we obtain a non-vanishing 
$\mbox{\boldmath$E$} \cdot \mbox{\boldmath$B$} / [B_{\rm eq}(2M)]^2$
at $t=0$ from the Wald solution,
as presented in the left panel of figure~\ref{fig:Epara_1}.
We also plot equi-$A_\varphi$ contours,
which represent the magnetic field lines in the poloidal plane
in a stationary and axisymmetric magnetosphere
(as in the case of the Wald solution),
as solid curves.
Here, $A_\varphi$ denotes the magnetic flux function,
and is the $\varphi$ component of the vector potential $A_\mu$,
which is related to the Faraday tensor by
$F_{\mu \nu} \equiv \partial_\mu A_\nu - \partial_\nu A_\mu$.

Since the contour interval is taken to be constant,
the density of the solid curves shows the magnetic-field strength.
In the right panel, we plot the magnetic field lines
with smaller contour interval than the left panel,
closing up the BH vicinity, $\varpi < 2M$,
where $\varpi$ (i.e., the abscissa) denotes the distance 
from the rotation axis measured in the Boyer-Lindquist $r$ coordinate.

We define that the magnetic field direction so that it may point upward 
in both hemispheres.
Thus, the yellow-red (or green-violet) region indicates
inward electric field in the upper (or lower) hemisphere.
This non-vanishing, magnetic-field-aligned electric field, 
will accelerate electrons outwards (or positrons inwards)
in the higher latitudes,
and accelerate electrons inwards (or positrons outwards)
in the lower latitudes.
The motion of such charges induces electric currents 
in the magnetosphere once the simulation begins at $t=0$.

\subsection{Electric fields and currents}
\label{sec:E_field}
\begin{figure*}
%\includegraphics[width=\columnwidth]{fig3.eps}
%\vspace*{-16.0truecm}
% \hspace{1.5cm}
\includegraphics[width=0.75\textwidth, angle=0]
{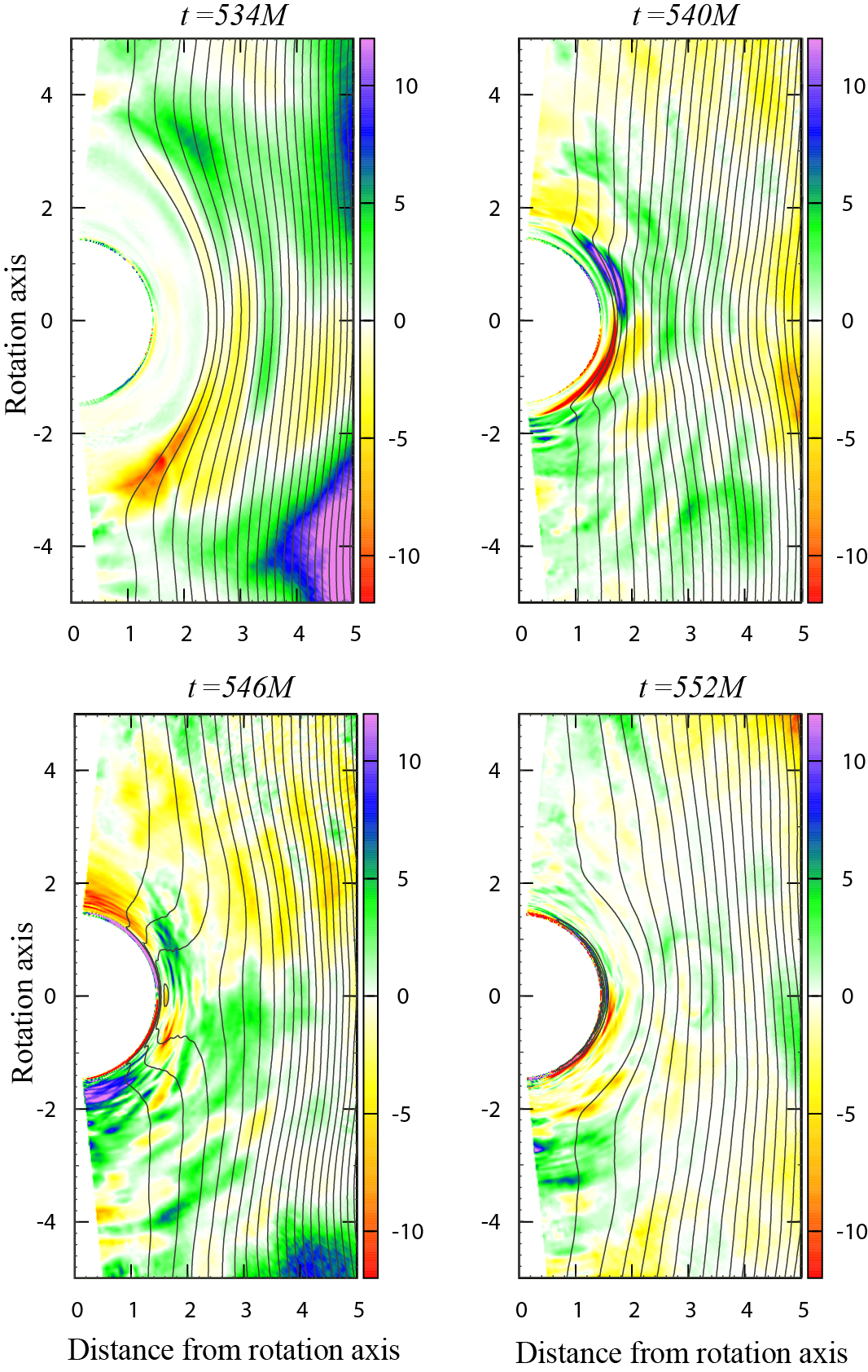}
% \vspace*{-0.0truecm}
% \includegraphics[width=\textwidth]{fig3.pdf}
% \vspace*{-6.0truecm}
\caption{
Magnetic-field-aligned electric field,
$\mbox{\boldmath$E$} \cdot \mbox{\boldmath$B$} 
 / B_{\rm eq}(2M){}^2$ 
(color image),
and the magnetic field lines (solid curves)
at four discrete elapsed times 
when $M=10M_\odot$, $a=0.9M$, and 
$\dot{m}=2.5 \times 10^{-4}$.
The horizon resides at $r= 1.435 M$
(although it is not depicted for clarity).
            }
    \label{fig:Epara_2}
\end{figure*}

As time elapses, the electric current carried by the charged leptons
alter the electric field through the Ampere's law,
and hence the magnetic field through the Faraday's law.
Accordingly, the acceleration electric field,
$\mbox{\boldmath$E$} \cdot \mbox{\boldmath$B$}/B$,
evolves to a qualitatively different configuration 
from the initial one.
Figure~\ref{fig:Epara_2}
shows 
$ \mbox{\boldmath$E$} \cdot \mbox{\boldmath$B$} / 
  [B_{\rm eq}(2M)]^2$
by the color images at four discrete elapsed times,
$t=534M$, $540M$, $546M$, and $552M$.
We also plot equi-$A_\varphi$ contours by black solid curves.
It follows that a non-vanishing
$\mbox{\boldmath$E$} \cdot \mbox{\boldmath$B$}$ 
changes with time,
and becomes much greater than $B_{\rm eq}{}^2$ 
in some limited regions.
At a small accretion rate $\dot{m}=2.5 \times 10^{-4}$,
the ADAF cannot supply enough MeV photons that 
are capable of materializing as pairs,
and cannot sustain the magnetosphere force-free.
We also find that
$\mbox{\boldmath$E$} \cdot \mbox{\boldmath$B$}$ 
peaks around the inner light surface
\citep[fig.~2 of][]{hiro16a} at $t=540M$, 
and that the magnetic field lines densely reside near the horizon
at the same timing, 
as the top right panel shows.
This point will be examined more closely 
in \S~\ref{sec:BZ_flux}
in relation to the energy extraction from the BH.
In the top left panel, magnetic field lines do not seem
to exist near the horizon;
however, it merely shows that the magnetic field is week there.

%Fig. 1

\begin{figure*}
%\includegraphics[width=\columnwidth]{fig3.eps}
%\vspace*{-16.0truecm}
\hspace{1.5cm}
\includegraphics[width=0.9\textwidth, angle=0]{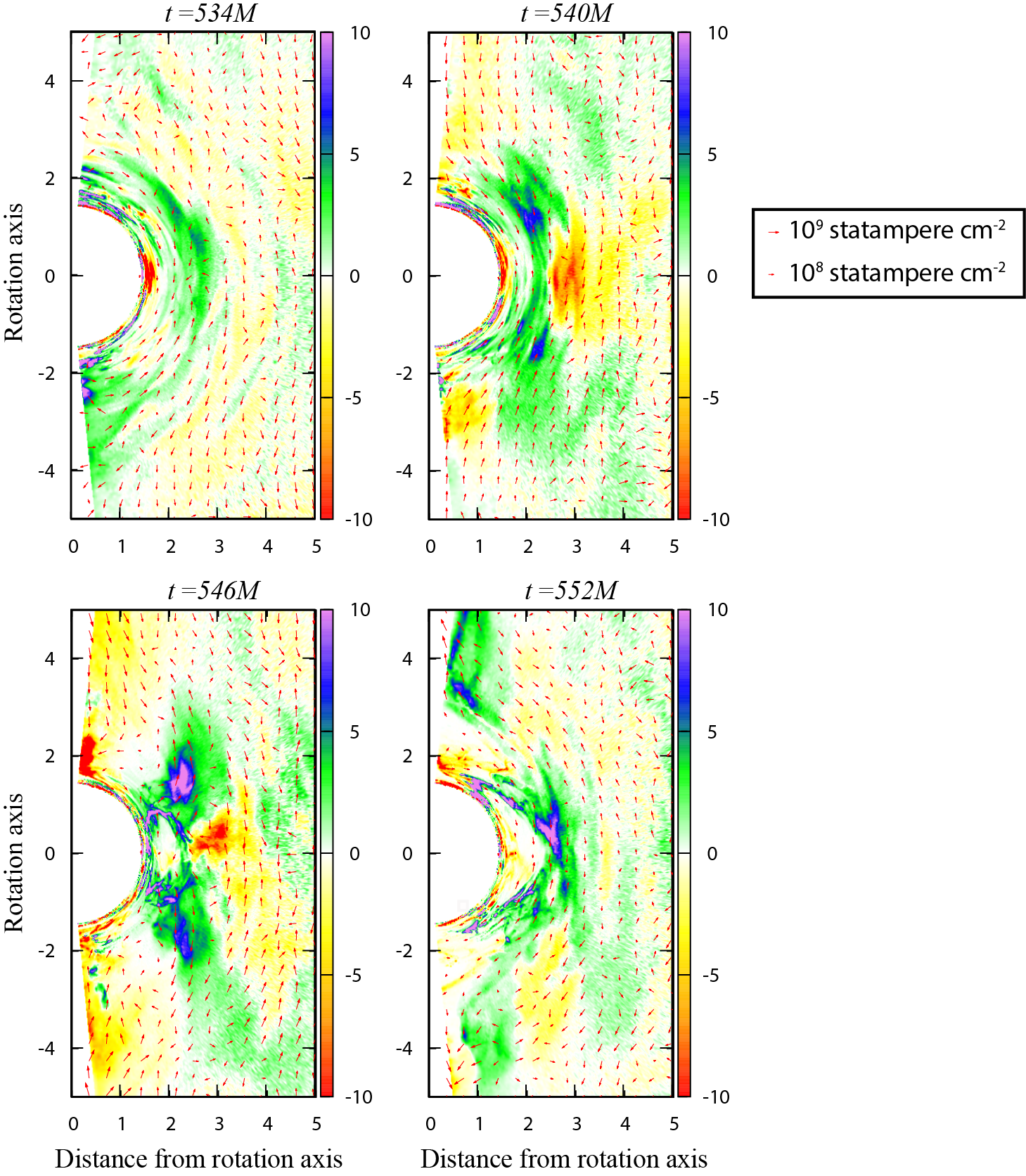}
\vspace*{-0.0truecm}
%\includegraphics[width=\textwidth]{fig3.pdf}
% \vspace*{-6.0truecm}
\caption{
Snap shots of the real charge density,
$(n_+ - n_-)/n_{\rm GJ}$ (color image), 
and electric currents (red arrows) 
at four discrete elapsed times as indicated.
The charge density is normalized by the Goldreich-Julian value,
and is depicted in linear scale as shown by the vertical color bars.
For instance, the green-violet (or yellow-red) regions 
show positive (or negative) 
dimensionless charge densities.
For presentation purpose, 
the currents are measured in ZAMO and depicted in cgs unit 
(i.e., in $\mbox{statampere cm}^{-2}$)
in logarithmic scale; 
example arrows are depicted in the right-most panel.
For presentation purpose,
we uniformly divide the poloidal plane 
in $X=r\sin\theta$ (abscissa) and $Y=r\cos\theta$ (ordinate) directions,
average the currents in each square cell,
and plot the averaged current by a red arrow 
from the center of each cell.
Both abscissa and ordinates are in $r_{\rm g}= M= GM c^{-2}$ unit.
The horizon resides at $r= 1.435 M$.
            }
    \label{fig:chJp}
\end{figure*}

In figure~\ref{fig:chJp},
we plot the real charge density (color image)
and the electric currents (red arrow) on the poloidal plane
at the same four discrete timings.
It shows that the charge density is comparable to
the Goldreich-Julian value.
Both the charge density and the current evolve
within several dynamical timescales.
It also follows from the top right panel
that a return current is formed towards the equator 
within the ergosphere at $t=540M$.
We will examine this point more closely
in \S~\ref{sec:BZ_flux}
in relation to the energy extraction from the BH.

\subsection{Magnetic field}
\label{sec:B_field}
\begin{figure*}
%\includegraphics[width=\columnwidth]{fig3.eps}
%\vspace*{-16.0truecm}
\hspace{1.5cm}
\includegraphics[width=0.8\textwidth, angle=0]{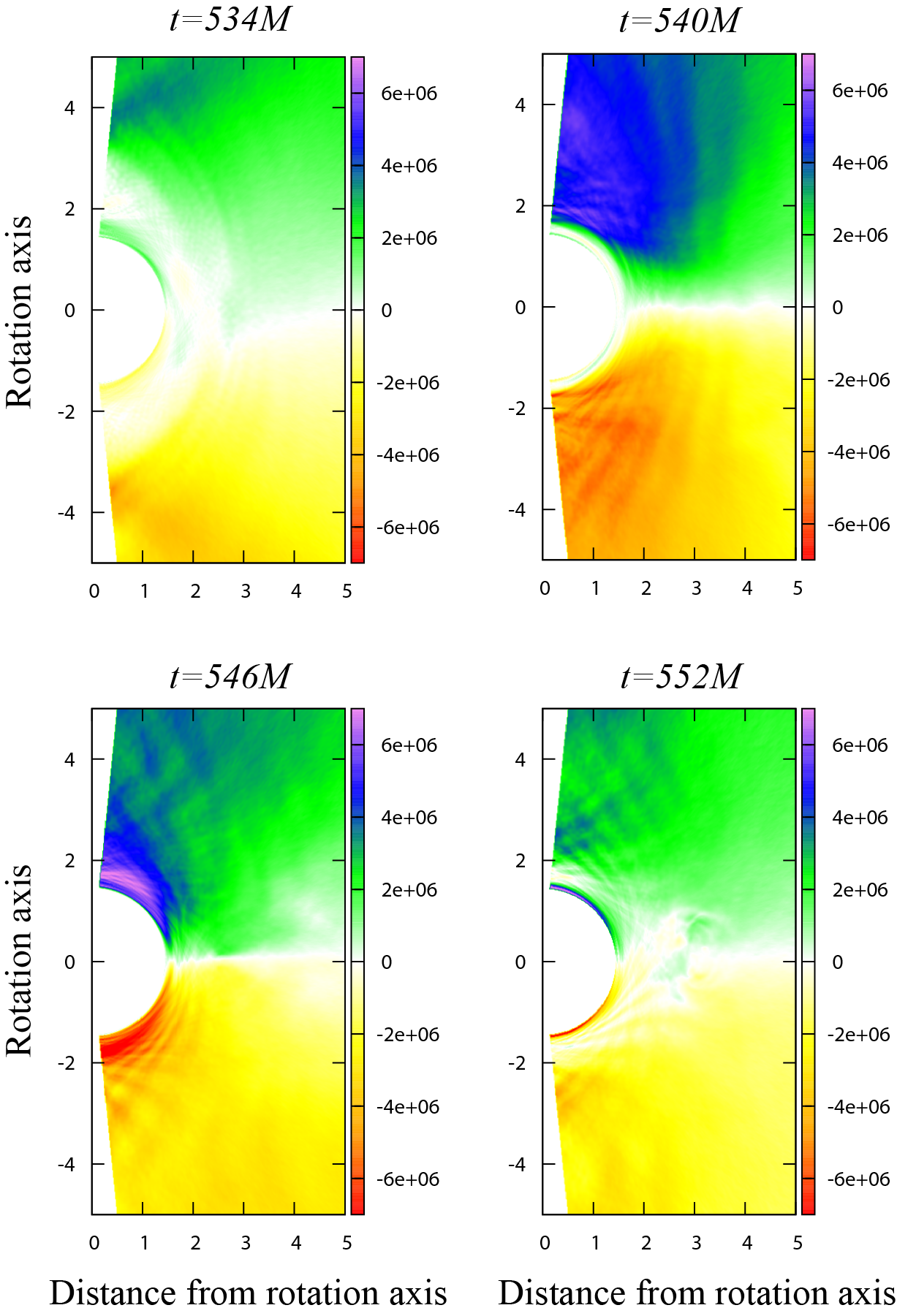}
\vspace*{-0.0truecm}
%\includegraphics[width=\textwidth]{fig3.pdf}
% \vspace*{-6.0truecm}
\caption{
Snap shots of radial component of the magnetic field measured by
Zero Angular Momentum Observer (ZAMO).
The color code shows its strength in gauss,
and is common in all the panels.
            }
    \label{fig:B1_4timings}
\end{figure*}

Let us next examine the magnetic field.
First, we plot the ZAMO-measured radial component
$B^{\hat r}= F_{\theta \varphi} / ( \sqrt{A} \sin\theta )$
in figure~\ref{fig:B1_4timings}.
It follows that $B^{\hat r}$ takes a large value 
in the higher and middle latitudes at $t \sim 540M$
and gradually decreases with time.
By symmetry, $B^{\hat r}$ changes sign between
the upper and the lower hemispheres.

\begin{figure*}
%\includegraphics[width=\columnwidth]{fig3.eps}
%\vspace*{-16.0truecm}
\hspace{1.5cm}
\includegraphics[width=0.8\textwidth, angle=0]{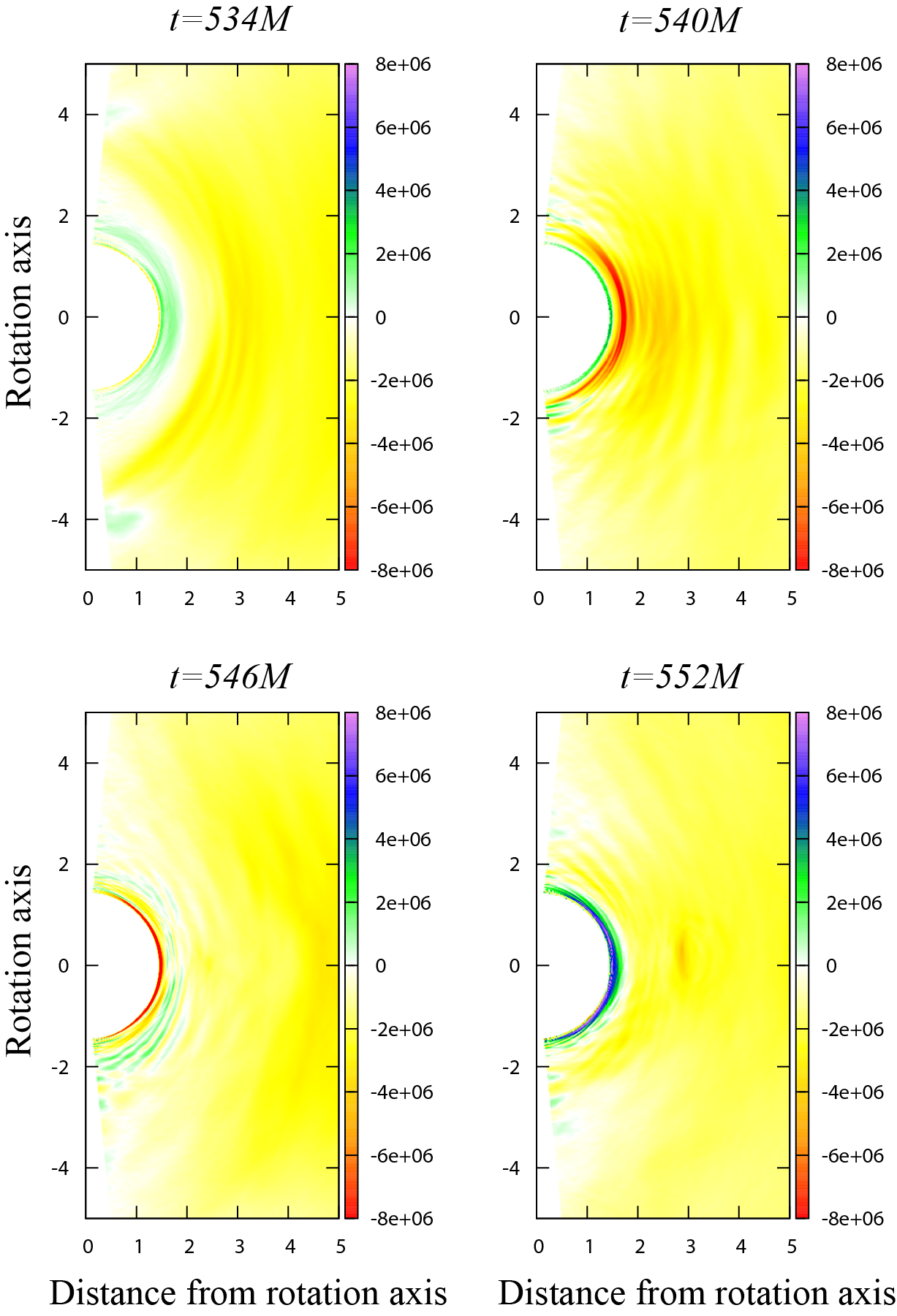}
\vspace*{-0.0truecm}
%\includegraphics[width=\textwidth]{fig3.pdf}
% \vspace*{-6.0truecm}
\caption{
Snap shots of the meridional component 
of the magnetic field measured by ZAMO.
The color code shows its strength in gauss.
            }
    \label{fig:B2_4timings}
\end{figure*}

Second, we plot the ZAMO-measured meridional component, 
$B^{\hat \theta}= \sqrt{\Delta/A} F_{\varphi r}/\sin\theta$, 
in figure~\ref{fig:B2_4timings}.
Except for the horizon vicinity, $B^{\hat \theta}$
takes negative values 
with smaller amplitude than $B^{\hat r}$.
We can also find large-amplitude oscillations 
appearing in radial direction
in the direct vicinity of the horizon.
This is due to the redshift effect
and will be discussed again in the final part of this
subsection.

\begin{figure*}
%\includegraphics[width=\columnwidth]{fig3.eps}
%\vspace*{-16.0truecm}
\hspace{1.5cm}
\includegraphics[width=0.8\textwidth, angle=0]{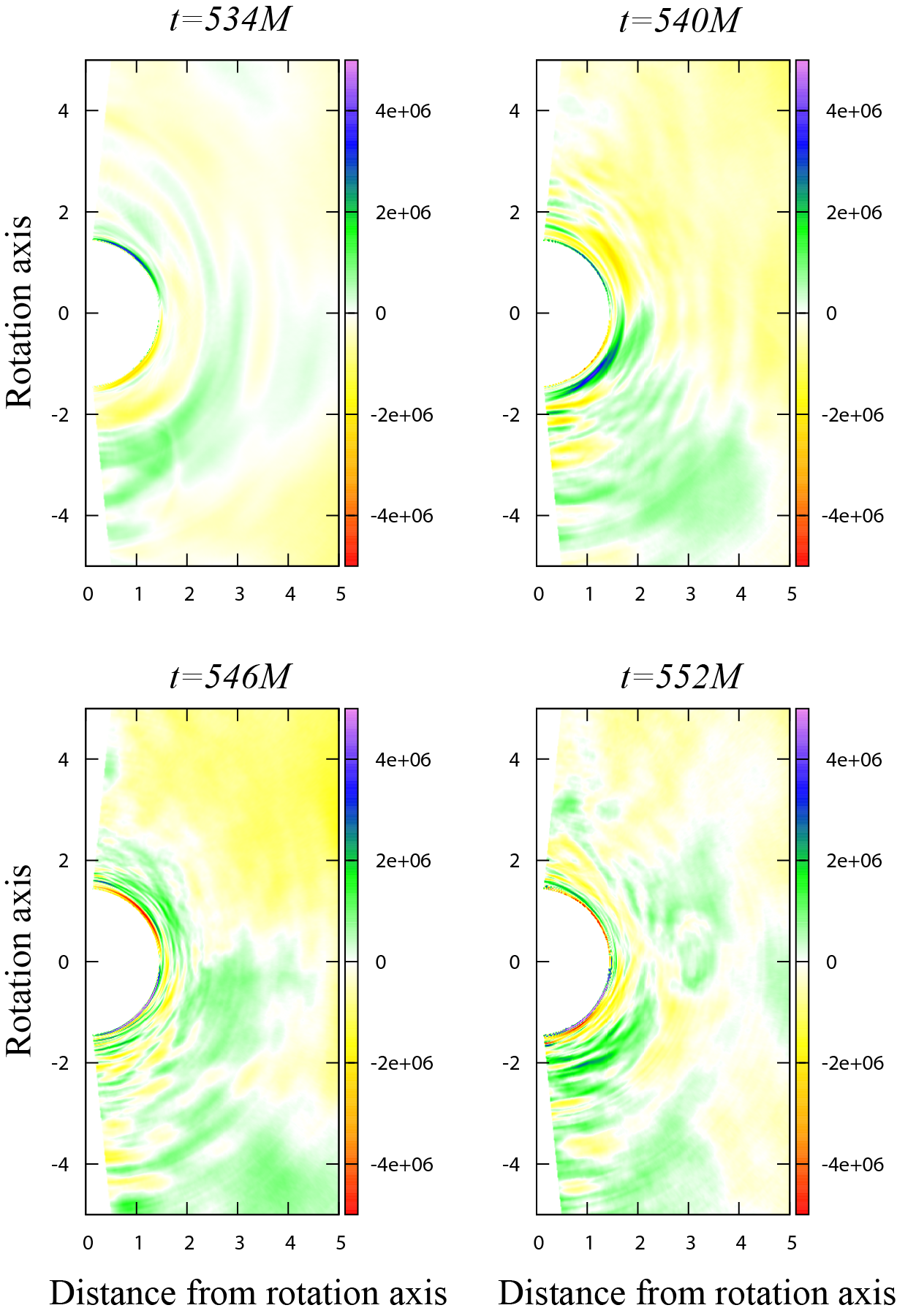}
\vspace*{-0.0truecm}
%\includegraphics[width=\textwidth]{fig3.pdf}
% \vspace*{-6.0truecm}
\caption{
Snap shots of toroidal component of the magnetic field measured by ZAMO.
The color code shows its strength in gauss.
            }
    \label{fig:B3_4timings}
\end{figure*}

Third, we plot the ZAMO-measured toroidal component,
$B^{\hat \varphi}= B_3 / \sqrt{\Delta}$
in figure~\ref{fig:B3_4timings}.
For this horizontal component at the horizon,
there appears a divergent factor, $\Delta^{-1/2}$, in the right-hand side,
because the ZAMO becomes an unphysical observer at the horizon.
(Note that we use ZAMO-measured quantities only for presentation purpose.
 We never adopt the ZAMO in any actual computation in our PIC scheme.)
Neglecting this artificial divergence 
(i.e., the ill behaviour of ZAMO) at the horizon,
we find $\vert B^{\hat \varphi} \vert < \vert B^{\hat r} \vert $
in most regions.
It is also clear that $B^{\hat \varphi}$ generally becomes negative
(or positive) in the upper (or lower) hemisphere,
because the magnetic field lines tend to be swept back by rotation.

\begin{figure*}
%\vspace*{-16.0truecm}
%\begin{minipage}[8.0cm]
%\hspace{2.3cm}
%\begin{minipage}[h]{0.5\textwidth}
\includegraphics[width=\textwidth]{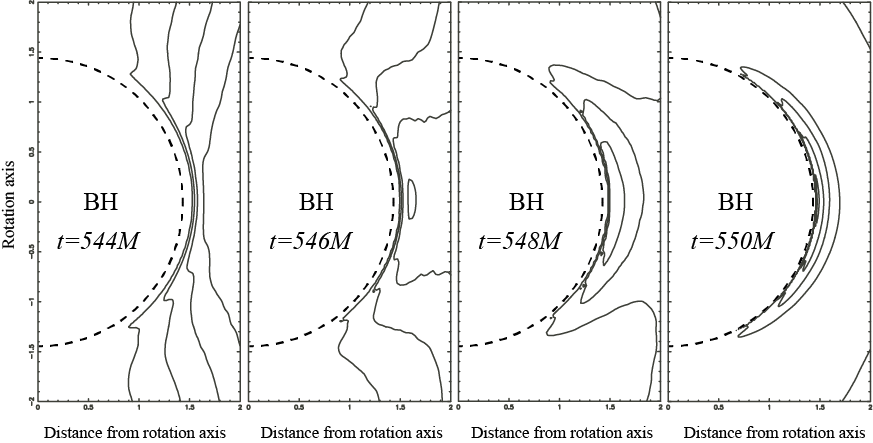}
%\vspace*{-1.5truecm}
%\end{minipage}
%\begin{minipage}[4.0cm]
%\end{minipage}
%\vspace{-2.0cm}
%\begin{minipage}[h]{0.3\textwidth}
\caption{
Close up of the equi-$A_\varphi$ contours (solid curves)
near the horizon at four discrete timings.
Both the abscissa and the ordinate designate
the Boyer-Lindquist $r$ coordinate in $r_{\rm g}=M$ unit.
The horizon resides at $r= 1.435M$ (on the dashed semicircle).
    }
        \label{fig:Blines}
%    \end{minipage}
\end{figure*}

Let us briefly examine the evolution of magnetic-field lines.
In figure~\ref{fig:Blines},
we present the equi-$A_\varphi$ contours near the horizon
at four discrete timings as indicated.
It is found that magnetic islands appear 
due to magnetic reconnection within the ergosphere
at $t \sim 546M$,
and migrate towards the BH, being elongated 
along the horizon due to the gravitational redshift,
as the right two panels show.
Because of this effect, $B^{\hat \theta}$ alternate the sign 
in the direct vicinity of the horizon,
as figure~\ref{fig:B2_4timings} demonstrates.
Note that it takes infinite time for magnetic islands to arrive
the horizon in the Boyer-Lindquist coordinates.
We thus adopt the tortoise coordinate, $r_\ast$, 
to resolve the anti-parallel magnetic field lines
accumulated in the direct vicinity of the horizon.

%Fig. 2

\subsection{Magnetic reconnection near the horison}
\label{sec:reconnection}
It is noteworthy that \citet{Crinquand:2021:A&A}
reported magnetic reconnections taking place on the equator,
including outside the static limit
(i.e., at $r>2M$ on the equator).
% \textbf{
In their simulation, their initial poloidal magnetic field
(their eq.~[9]) quickly dies out in a few tens of $M=r_{\rm g}/c$.
Accordingly, relatively weak poloidal magnetic field outside
the ergosphere allowed X and O points 
(where the magnetic field forms an X-like or O-like geometry
 in the poloidal plane)
to arise at $r>3M$ (their fig.~4).
On the other hand, in the present analysis,
we superpose stationary Wald electromagnetic fields
on non-stationary PIC fields,
assuming a stationary ring current 
at some large-enough distance on the equator.
As a result, the initial poloidal magnetic field does not 
die out and prevents X and O points to arise at $r>2M$.
That is, magnetic reconnection takes place efficiently only 
inside the static limit.
% }

\begin{figure*}
\begin{interactive}{animation}{Aph_Npe_515.gif}
\plotone{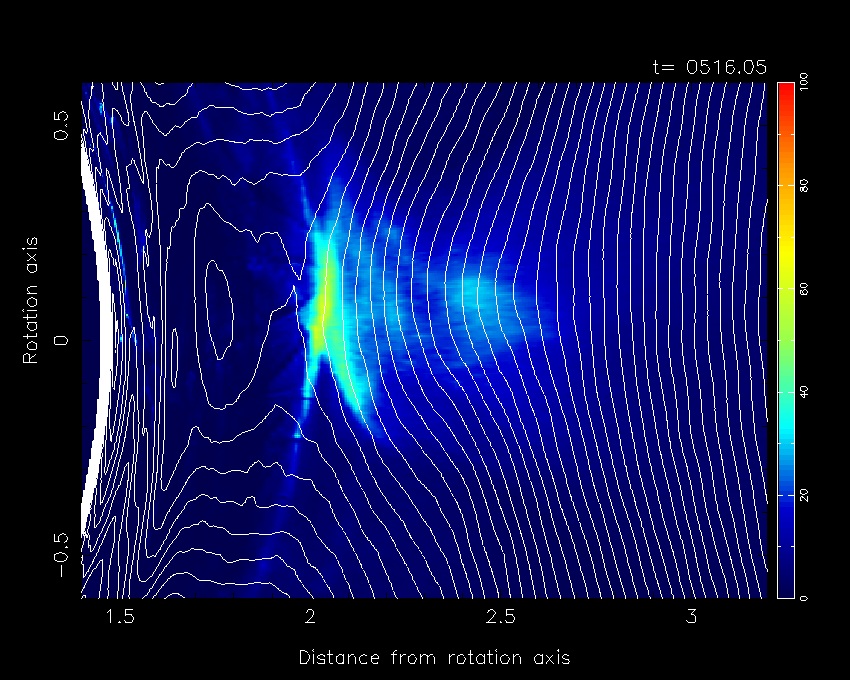}
\end{interactive}
\caption{
% \textbf{
Magnetic field lines (white curves) and 
the pair density (color, in Goldreich-Julian unit)
on the poloidal plane near the horizon,
$1.4M < x= r \sin\theta < 3.2M$ and
$\vert r \cos\theta \vert < 0.6M$,
at elapsed time $t=516.05M$.
The ordinate $y= r\cos\theta=0$ corresponds to the equatorial plane.
Magnetic reconnection takes place at the X-type point
that appears at $(x,y)=(1.95M,0.14M)$
at this timing.
The horizon resides at $r= 1.435M$ in the left-most part.
An animation of this simulation is available in the online journal. 
The animation covers the simulation from $t= 515.60M$ to $516.99M$.
% }
\label{fig:movie_reconnection}}
\end{figure*}

% \textbf{
In figure~\ref{fig:movie_reconnection},
we plot the equi-$A_\varphi$ contours by white curves
and the total pair number density, $(n_- + n_+)/n_{\rm GJ}$ in color.
It follows from the online animation 
that magnetic reconnections subsequently take place
at the X-type point, 
which is located at $r < 2M$ and $\theta \sim \pi/2$
(i.e., inside the ergosphere near the equator).  
Pair plasmas are ejected from this reconnection region horizontally 
both rightward and leftward.
Plasmas ejected rightward (i.e., into greater $r$) are compressed
outside the ergosphere, $2.0 M < r < 2.1 M$.
On the other hand, 
plasmas ejected leftward (i.e., toward the horizon) are not efficiently
compressed around the O-type point (located around $1.7M < r < 1.8M$),
because the magnetic-field strength is weak there
compared to those at $r > 2 M$, 
as can be seen from the difference of contour intervals.
Instead, in-falling plasmas are compressed in the direct vicinity 
of the horizon (in the Boyer-Lindquist coordinates);
however, the resultant high-density regions (in red color)
are hidden under the piled-up magnetic field lines, 
which appear as a bunch of white curves right above the horizon.
% }

% \begin{figure*}
% \includegraphics[width=\textwidth]{f_npe_Aph_m25_5452_global.png}
% \caption{
% Total particle number density, $(n_+ + n_-) / n_{\rm GJ}$,
% (color) and  equi-$A_\varphi$ contours (solid curves)
% in the equatorial region
% at $t=520.20M$.
% Both the abscissa and the ordinate designate
% the Boyer-Lindquist $r$ coordinate in $r_{\rm g}=M$ unit.
% The horizon resides at $r= 1.435M$.
%      }
% \label{fig:n_tot_global}
% \end{figure*}
% 
% \begin{figure*}
% \includegraphics[width=\textwidth]{f_npe_Aph_m25_5452.png}
% \caption{
% Same figure as fig.~\ref{fig:n_tot_global},
% but the inner region is magnified. 
%      }
% \label{fig:n_tot}
% \end{figure*}
% 
% \begin{figure*}
% \includegraphics[width=\textwidth]{f_B2E2_Aph_m25_5452.png}
% \caption{
% Similar figure as fig.~\ref{fig:n_tot_global},
% but $(B^2-E^2) / B_{\rm eq}{}^2$ is plotted
% at t=520.20M. 
%      }
% \label{fig:B2E2}
% \end{figure*}
% 
% \begin{figure*}
% \includegraphics[width=\textwidth]{f_F30_Aph_m25_5452.png}
% \caption{
% Similar figure as fig.~\ref{fig:n_tot_global},
% but $F_{\varphi t} / B_{\rm eq}$ is plotted
% at t=520.20M. 
%      }
% \label{fig:F30}
% \end{figure*}

\subsection{Particle energy distribution}
\label{sec:particle_energy}
\begin{figure*}
%\includegraphics[width=\columnwidth]{fig3.eps}
%\vspace*{-16.0truecm}
% \hspace{1.5cm}
\includegraphics[width=\textwidth, angle=0]{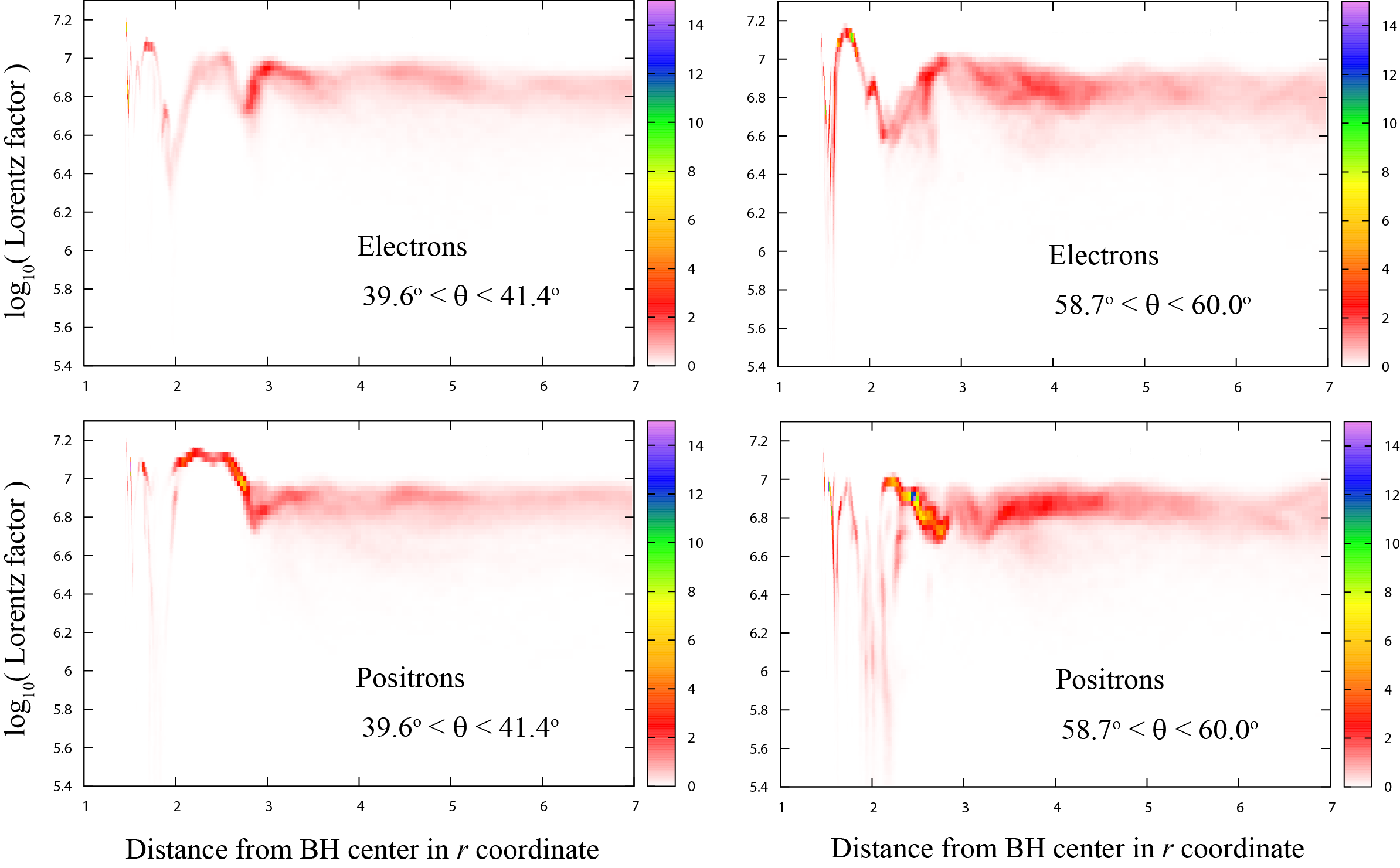}
% \vspace*{-0.0truecm}
%\includegraphics[width=\textwidth]{fig3.pdf}
% \vspace*{-6.0truecm}
\caption{
Energy distribution of electrons (top panels) and positrons (bottom panels)
within the indicated colatitude.
The abscissa refers to the distance from the BH center 
in the Boyer-Lindquist radial coordinate.
            }
    \label{fig:ptcl_Lf}
\end{figure*}

Let us briefly browse the energy dependence of macro particles.
In figure~\ref{fig:ptcl_Lf},
we plot the distribution of 
$u^t = u^0= dt/d\tau$
as a function of the distance from the BH.
In special relativity, $u^t$ refers to the Lorentz factor.
However, in the present case, $u^t$ also contains the time dilation
due to the gravitational redshift and the frame dragging
at the particle's position.
Nevertheless, for simplicity, we represent $u^t$ as
the \lq Lorentz factor' in this figure.
The color image shows the particle distribution function
normalized by the Goldreich-Julian number density.
For the details of this normalization, see \S~3.5.2 of Paper~I,
where $\gamma$ in equation~(60) of Paper~I corresponds to $u^t$
in the present notation.

We plot the electron (or positron) energy distributions at $t=540M$
as the top (or bottom) two panels.
The left two panels show the distribution within the colatitude
$39.69^\circ < \theta < 41.42^\circ$,
and the right ones do within 
$58.72^\circ < \theta < 60.02^\circ$.
It is found that the 
particle Lorentz factors are kept above $2 \times 10^5$ at each position.
Because a strong acceleration electric field arises 
near the inner light surface at $t \sim 540M$,
the Lorentz factor distribution peaks within $r < 1.8M$ 
for both electrons and positrons at this timing, $t=540M$.
If we integrate over the particles within the specified 
latitudinal range at each $r$ at elapsed time $t=540M$,
we find that the positronic (or electronic) density peaks
at $\sim 36 n_{\rm GJ}$ (or $\sim 30 n_{\rm GJ}$) 
with averaged Lorentz factor
$\langle \gamma \rangle \sim 8.3 \times 10^6$ 
(or $\sim 6.3 \times 10^6$) 
in the region
$2.4M < r < 2.5M$ and $50^\circ < \theta < 70^\circ$.

\subsection{The Blandford-Znajek flux}
\label{sec:BZ_flux}
Now let us consider the BZ flux.
The radial component of the BZ flux 
(i.e., the Poynting flux) become,
\begin{eqnarray}
  \lefteqn{T_{\rm em}{}^r{}_t 
           = \frac{c}{4\pi} F^{r \mu} F_{\mu t}
          }
    \nonumber\\
    &=& 
    \frac{c}{4\pi}\frac{1}{\Sigma}
    \nonumber\\
    &\times&
    \left[ B_3 E_2
              - \left( \frac{2Mar}{\Sigma} E_1
                      +\frac{\Delta-\Sigma a^2\sin^2\theta}
                            {\Sigma \sin^2\theta} B_2
               \right) E_3
        \right].
    \nonumber\\
   \label{eq:BZflux_r}
\end{eqnarray}
% \begin{eqnarray}
%   \lefteqn{T_{\rm em}{}^\theta{}_t 
%            = \frac{c}{4\pi} F^{\theta \mu} F_{\mu t}
%           }
%     \nonumber\\
%     &=& 
%     \frac{c}{4\pi}\frac{1}{\Sigma}
%     \nonumber\\
%     &\times&
%     \left[ -B_3 E_1
%               + \left( \frac{2Mar}{\Sigma} E_2
%                       +\frac{\Delta-\Sigma a^2\sin^2\theta}
%                             {\Sigma \sin\theta} B_1
%                \right) \frac{E_3}{\Delta}
%         \right].
%     \nonumber\\
%   \label{eq:BZflux_th}
% \end{eqnarray}
In what follows, we normalize 
$T_{\rm em}{}^r{}_t$ % and $T_{\rm em}{}^\theta{}_t$ 
with its analytically inferred value, 
$F_{\rm BZ}^{\rm ana}(r)$, where 
$F_{\rm BZ}^{\rm ana}(r) \equiv 
 L_{\rm BZ}^{\rm ana}/(4\pi r^2 f_{\rm corr}(r))$
denotes the typical BZ flux obtained by dividing 
the typical BZ luminosity $L_{\rm BZ}^{\rm ana}$ with 
the surface area $4 \pi r^2 f_{\rm corr}$,
and
$f_{\rm corr} \equiv 
  \int_0^1 \sqrt{(1+a^2/r^2)^2-(\Delta a^2/r^4) (1-z^2)} dz$.
The typical BZ luminosity is estimated by
\citep{Tchekhovskoy:2010:ApJ}
\begin{equation}
  L_{\rm BZ}^{\rm ana}
  = k \Phi^2 \frac{a^2}{16M^2},
  \label{eq:L_BZ_ana}
\end{equation}
where the total magnetic flux threading the horizon is given by
\begin{equation}
  \Phi= B_{\rm H} (4\pi M r_{\rm H})^2,  
\end{equation}
and $k \approx 1/6\pi$ for a radial magnetic field near the horizon.
The present magnetosphere is magnetically dominated,
in the sense that the magnetic energy density dominates
the particles' rest-mass energy densities.
Accordingly, particle contribution is negligible when we discuss the energy flux.

\begin{figure*}
%\hspace*{2.0truecm}
%\includegraphics[width=\columnwidth]{fig2.eps}
%\includegraphics[width=\textwidth]{fig3.pdf}
%\hspace{2.2cm}
\includegraphics[width=\textwidth, angle=0]
{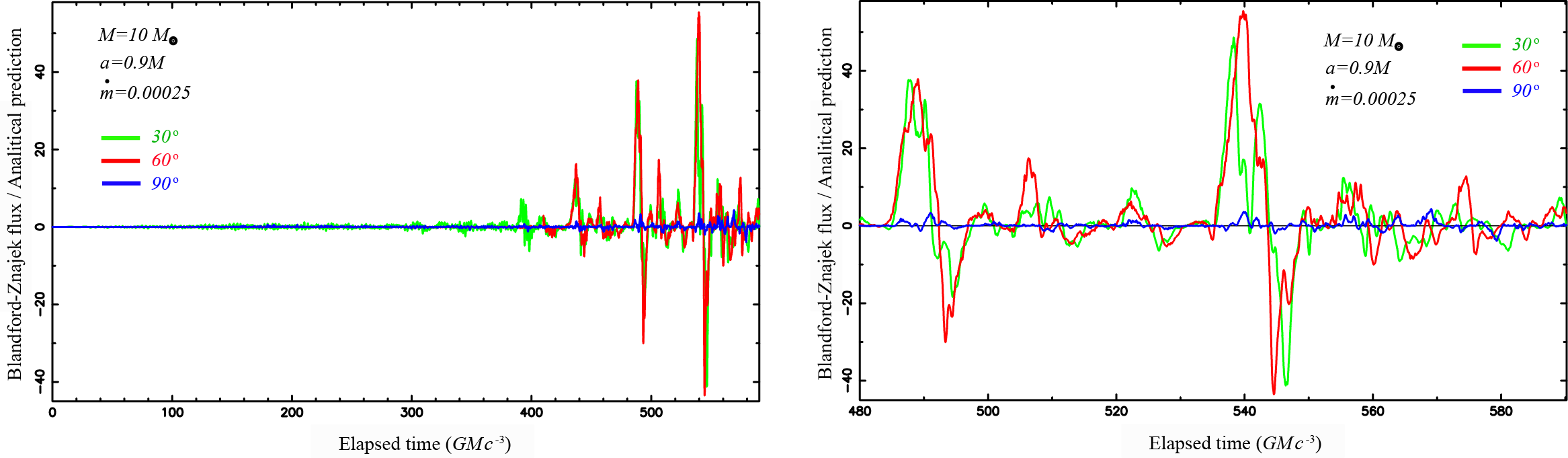}
\vspace*{-0.0truecm}
\caption{
Evolution of the Blandford-Znajek (BZ) flux 
at three discrete colatitudes in the upper hemisphere.
The abscissa is in the dynamical time scale ($M=GM c^{-3}$) unit,
while the ordinate is normalized by the analytical prediction
of the BZ flux.
The flux is measured at radius $r=2.335M$.
The left panel shows the BZ flux during
the entire simulation period,
while the right one shows its close-up at $t>480M$.
}
    \label{fig:LBZ_Time}
\end{figure*}

In figure~\ref{fig:LBZ_Time},
we plot the normalized BZ flux, 
$T_{\rm em}{}^r{}_t / F_{\rm BZ}{}^{\rm ana}$,
as a function of the dimensionless elapsed time $t/M$,
at position $r=2.335M$ and 
$\theta= 30^\circ$, $60^\circ$, and $90^\circ$.
The left panel shows the BZ fluxes during the entire simulation period,
while the right panel focuses the elapsed time after $t=480M$.
It follows that the flux oscillates between positive and negative
values and peaks with an interval $\sim 50M$.
This typical flaring period, $\sim 50M$,
does not change if we adopt different outer boundary radius,
such as $r_{\rm out}=20M$ or $r_{\rm out}=30M$, 
instead of $r_{\rm out}=25M$.

\begin{figure*}
%\hspace*{2.0truecm}
%\includegraphics[width=\columnwidth]{fig2.eps}
%\includegraphics[width=\textwidth]{fig3.pdf}
%\hspace{2.2cm}
\includegraphics[width=\textwidth, angle=0]{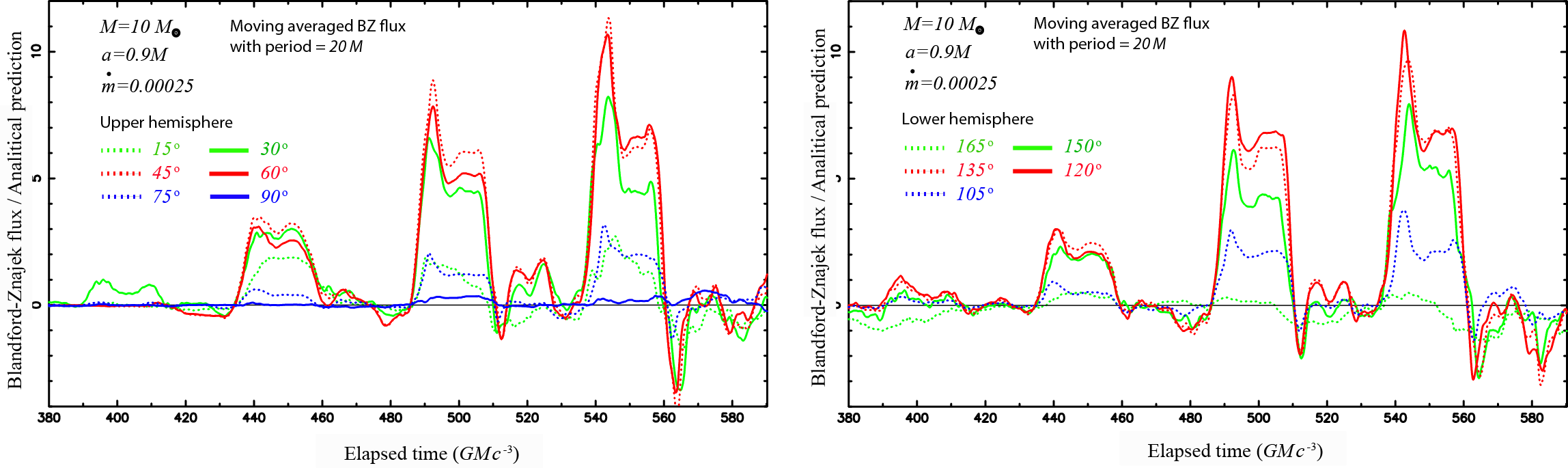}
\vspace*{-0.0truecm}
\caption{
Similar figure as fig.~\ref{fig:LBZ_Time} but moving-averaged
with a period 20M.
The flux is measured at radius $r=2.335M$.
The left panel shows the BZ fluxes in the upper hemisphere
at six discrete colatitudes, 
and the right panel does those in the lower hemisphere
at five colatitudes.
}
    \label{fig:LBZ_Time_MA}
\end{figure*}

To see the averaged flux, we take a moving average with
period $20M$.
Figure~\ref{fig:LBZ_Time_MA} shows the result
for the upper hemisphere (left panel) and 
the lower hemisphere (right panel).
It follows that the net flux is positive in both hemispheres,
which means that the BH's rotational energy is extracted
in the form of the Poynting flux,
and that the solution is more or less symmetric
between the upper and lower hemispheres.
We can also find that the BZ flux concentrates
in the middle latitude (red curves)
rather than in the higher (green) or lower (blue) latitudes.

It is worth considering the relation between the BZ-flux variation
and the magnetic-island evolution.
As figure~\ref{fig:Blines} indicates,
when a magnetic island is formed near the equator within the ergosphere, 
it expels magnetic field lines outside.
As time elapses, the magnetic island migrates inwards
being elongated along the horizon,
and eventually sticks to it and virtually disappear,
as the right-most panel of figure~\ref{fig:Blines} shows.
After the magnetic islands virtually disappear,
magnetic flux tubes return to the horizon vicinity,
as the top-right and bottom-left panels of figure~\ref{fig:Epara_2} show,
or the left-most panel of figure~\ref{fig:Blines} shows.
During this phase, magnetic field becomes strong
as the top-right panel of figure~\ref{fig:B1_4timings} shows,
inducing a strong acceleration electric field near the inner light surface,
as the top-right panel of figure~\ref{fig:Epara_2} shows.
It also follows from the top-right panel of figure~\ref{fig:chJp}
that a strong return current is formed at the same timing
within the ergosphere towards the equator.
This meridional current acts Lorentz forces on the magnetized plasmas 
pushing them into negative-energy orbits,
and exerts a counter torque on the horizon.
Accordingly, the BH's rotational energy is efficiently extracted.

On the other hand, when there exists a giant magnetic island 
within the ergosphere, the weak magnetic field near the horizon
cannot facilitate the BZ process efficiently.
Indeed, magnetic islands occupy a good fraction of the ergosphere
in a large fraction of time.
As a result, the BZ flux increases only sporadically,
as figure~\ref{fig:LBZ_Time} shows.
The characteristic flaring period, $\sim 50M$, 
is regulated by the time scale of reconnection
taking place within the ergosphere.

It is worth noting that the BZ process is facilitated
by virture of these return currents {\it outside} the horizon.
Assuming a stationary and axisymmetric BH magnetosphere,
\citet{Punsly:1996:ApJ} pointed out the infeasibility of the BZ process
in relation to the causality and the plasma's inertia.
However, this criticism can be overcome by the formation of 
these strong, time-dependent return currents within the ergosphere.

Since we are not considering the pair production
between the gap-emitted $\gamma$-rays and disk-emitted soft photons, 
we do not find the quasi-periodic gap activities 
through gap reopening and resultant pair cascade
as reported in 1D GRPIC simulations with photon tracking
\citep{Kisaka:2020:ApJ,Chen:2020:ApJ}.

\begin{figure*}
%\hspace*{2.0truecm}
%\includegraphics[width=\columnwidth]{fig2.eps}
%\includegraphics[width=\textwidth]{fig3.pdf}
%\hspace{2.2cm}
\includegraphics[width=\textwidth, angle=0]{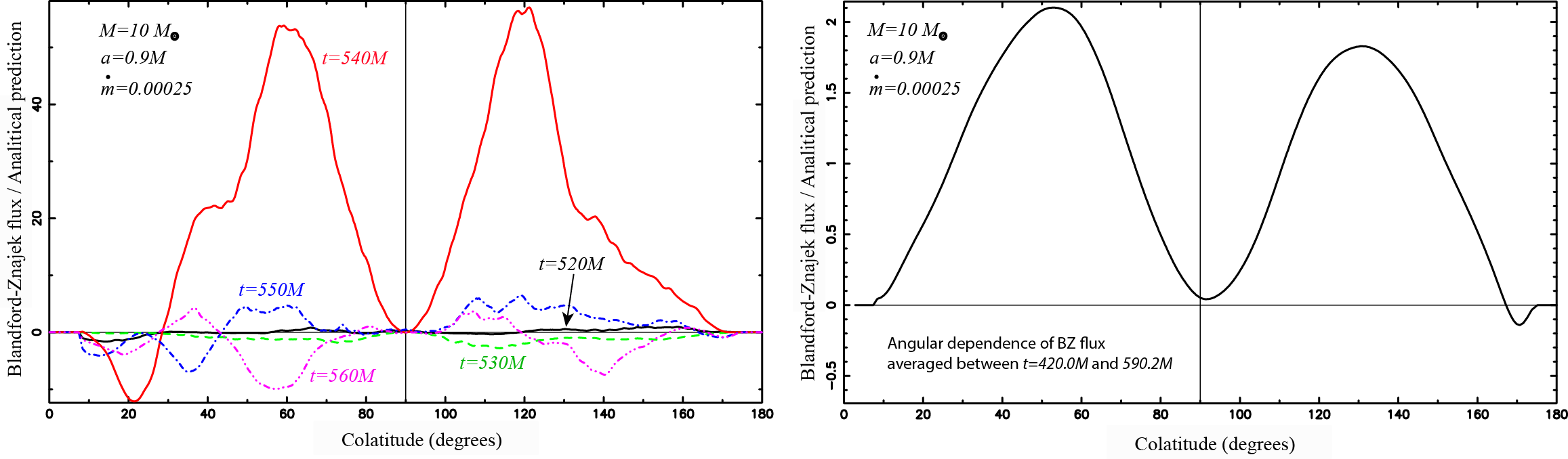}
\vspace*{-0.0truecm}
\caption{
Angular dependence of the BZ flux measured at radius $r=2.335M$.
{\it Left:} the BZ fluxes at five discrete elapsed times.
{\it Right:} the BZ flux averaged over time between $t=480M$ and $590M$.
The left (or right) half of each panel
corresponds to the upper (or lower) hemisphere.
}
    \label{fig:LBZ_angle}
\end{figure*}

To examine the angular dependence of the BZ flux,
we plot the BZ flux as a function of the colatitude
at five discrete elapsed times
in the left panel of figure~\ref{fig:LBZ_angle}.
At $t=540M$, the BZ flux peaks at 
$\theta \sim 60^\circ$ and $120^\circ$.
If we take a time average after $t=480M$,
we find that the BZ flux does peak in the middle latitude
in $40^\circ < \theta < 60^\circ$ (upper hemisphere) and
in $120^\circ < \theta < 140^\circ$ (lower hemisphere),
as the right panel indicates.

\begin{figure*}
%\hspace*{2.0truecm}
%\includegraphics[width=\columnwidth]{fig2.eps}
%\includegraphics[width=\textwidth]{fig3.pdf}
%\hspace{2.2cm}
\includegraphics[width=\textwidth, angle=0]{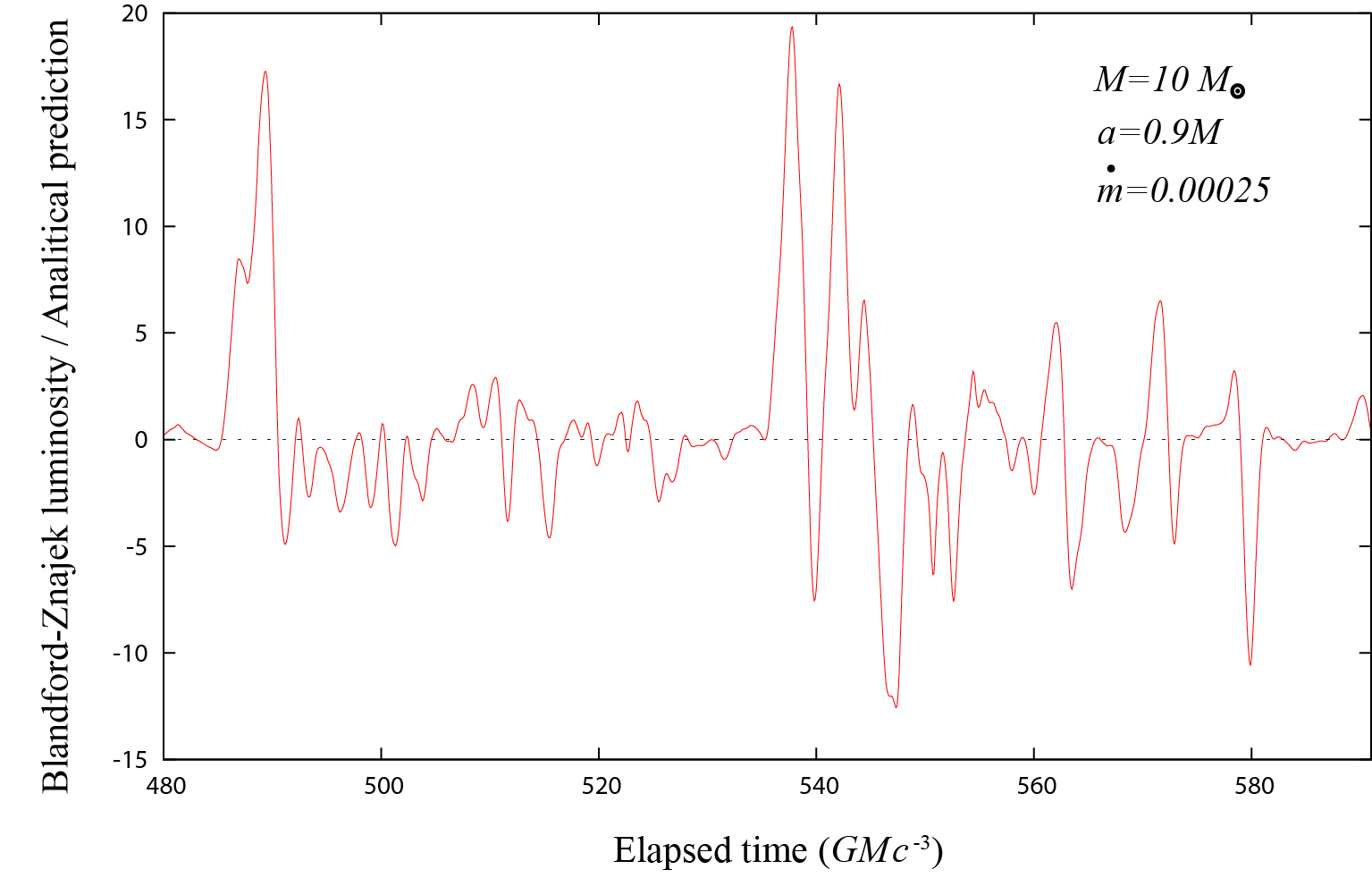}
\vspace*{-0.0truecm}
\caption{
Blandford-Znajek luminosity obtained by integrating its flux
over the entire spherical surface at $r=2.335M$.
The ordinate is normalized by the analytically inferred value.
}
    \label{fig:BZ_luminosity}
\end{figure*}

Let us quickly take a look at the BZ luminosity,
by integrating the BZ flux over the entire spherical surface 
at a fixed $r$.
When $\dot{m}=2.5 \times 10^{-4}$,
we can evaluate the magnetic-field strength at the horizon
by its equipartition value, $B_{\rm eq}=1.53 \times 10^6$~G at $r=2M$,
which allows us to analytically infer the luminosity as
$L_{\rm BZ}^{\rm ana}= 4.16 \times 10^{34} \mbox{ergs s}^{-1}$.
Evaluating the simulated flux at $r=2.335M$, 
and normalizing its integrated value with $L_{\rm BZ}^{\rm ana}$,
we obtain the BZ luminosity as presented 
in figure~\ref{fig:BZ_luminosity}.
We find that the BZ luminosity flares every $\sim 50$ dynamical time scales,
as expected, and the peak values attain $16-19$ times of the analytically
inferred value.

The time-averaged BZ luminosity is found to be
$2.63 \times 10^{33} \mbox{ergs s}^{-1}$,
which corresponds to $6.2$~\% of the analytical value.
Therefore, when the accretion rate is as small as 
$\dot{m}=2.5 \times 10^{-4}$,
we can conclude 
that the spin-down luminosity of the BH exceeds $15$ times
of the analytical estimate during the flare
although its long-term average is kept at only $6$~\% of it.

\subsection{Power spectrum of the BZ flux}
\label{sec:BZflux_PSD}
To look more closely at the time variability properties,
let us Fourier-transform the BZ flux.
For this purpose, we introduce a normalized 
power spectral density (PSD), $P(f_k)$, 
of function $f(t)= T_{\rm em}{}^r{}_t / F_{\rm BZ}{}^{\rm ana}$, 
so that the normalization may be defined by
\begin{equation}
  \frac{1}{T} 
  \sum_{k=0}^{N-1} \vert f(t_k) \vert^2 \Delta_t
  = \frac{1}{N} \sum_{k=0}^{N/2} P(f_k),
\end{equation}
where in the right-hand side
$f_k= k/(N \Delta_t)$ denotes the frequency. 
The function (in this particular case, the BZ flux) $f(t)$
is sampled at $N$ points, which span a range of time 
$T=(N-1) \Delta_t$.
The sampling interval is $\Delta_t= 1.098 \times 10^{-3} GMc^{-3}$;
thus, the Nyquist frequency is 
$f_{\rm c}= 4.796 \times 10^2 M^{-1}$.
% $f_{\rm c}= 4.79662774 \times 10^2 M^{-1}$.

\begin{figure*}
%\hspace*{2.0truecm}
%\includegraphics[width=\columnwidth]{fig2.eps}
%\includegraphics[width=\textwidth]{fig3.pdf}
%\hspace{2.2cm}
\includegraphics[width=\textwidth, angle=0]{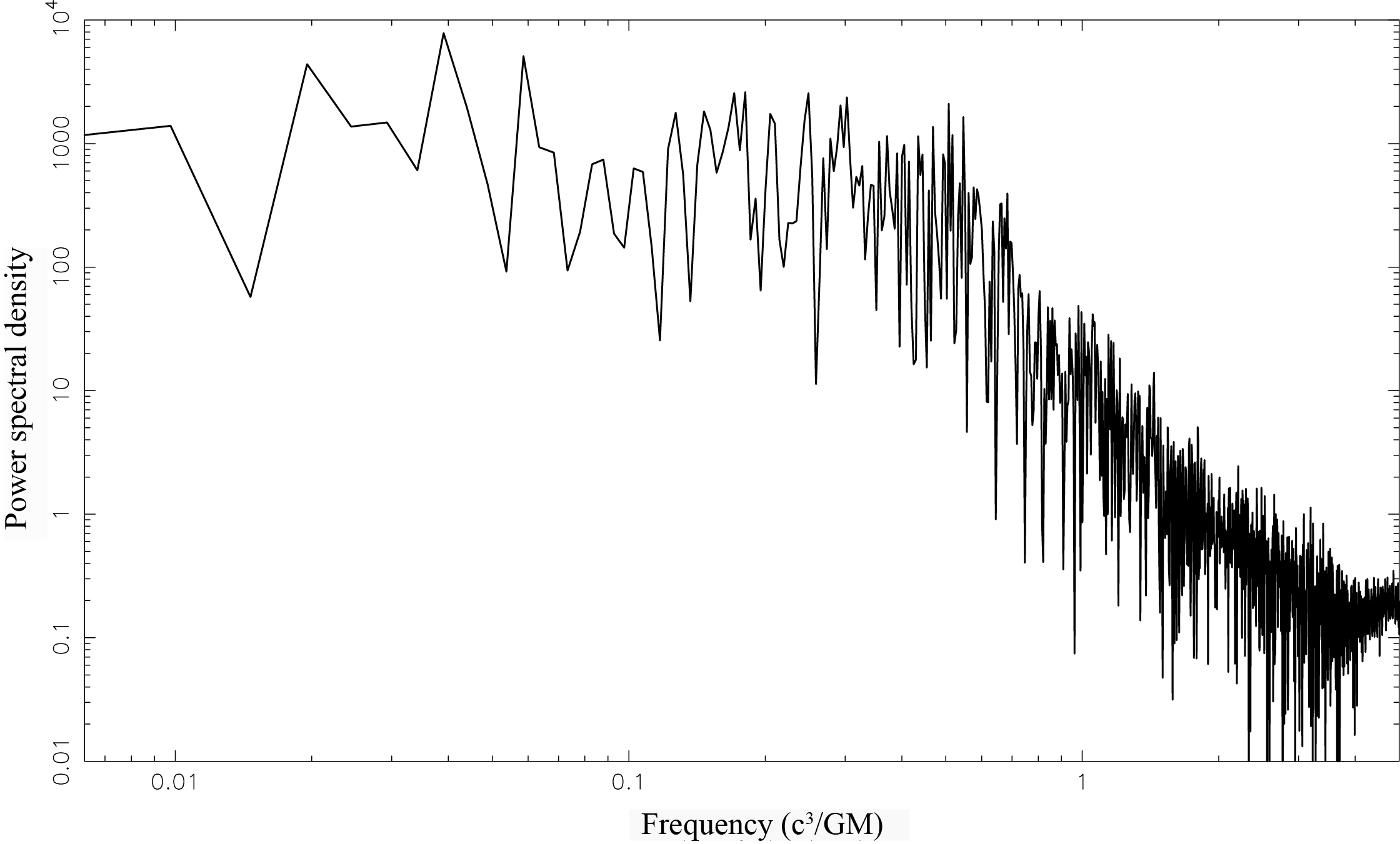}
\vspace*{-0.0truecm}
\caption{
Normalized power spectral density of the Blandford-Znajek flux
at colatitude $\theta= 45^\circ$ and radius $r=2.335M$.
}
    \label{fig:LBZ_PSD}
\end{figure*}

We plot the PSD  % and \ref{fig:LBZ_PSD2},
at $\theta=45^\circ$ in figure~\ref{fig:LBZ_PSD}.
It follows that a quasi-periodic oscillation (QPO) appears
around $0.02 M^{-1}$,
and its higher harmonics appear around
$0.04 M^{-1}$ and $0.06 M^{-1}$.
% , and around $0.0240 M^{-1}$
% for $\dot{m}=3 \times 10^{-4}$ (fig.~\ref{fig:LBZ_PSD2}).
The fundamental frequency of the QPO, $\sim 0.02M^{-1}$, 
means that 
there is a modulation of the amplitude with a period 
$P_{\rm QPO} \sim 50 M$,
which corresponds to interval of the flares
that can be seen in figures~\ref{fig:LBZ_Time} and \ref{fig:LBZ_Time_MA}.

%Fig. 5

% \textbf{
\section{The case of greater accretion rate}
\label{sec:larger_mdot}
\begin{figure*}
%\hspace*{2.0truecm}
%\includegraphics[width=\columnwidth]{fig2.eps}
%\includegraphics[width=\textwidth]{fig3.pdf}
%\hspace{2.2cm}
\includegraphics[width=\textwidth, angle=0]{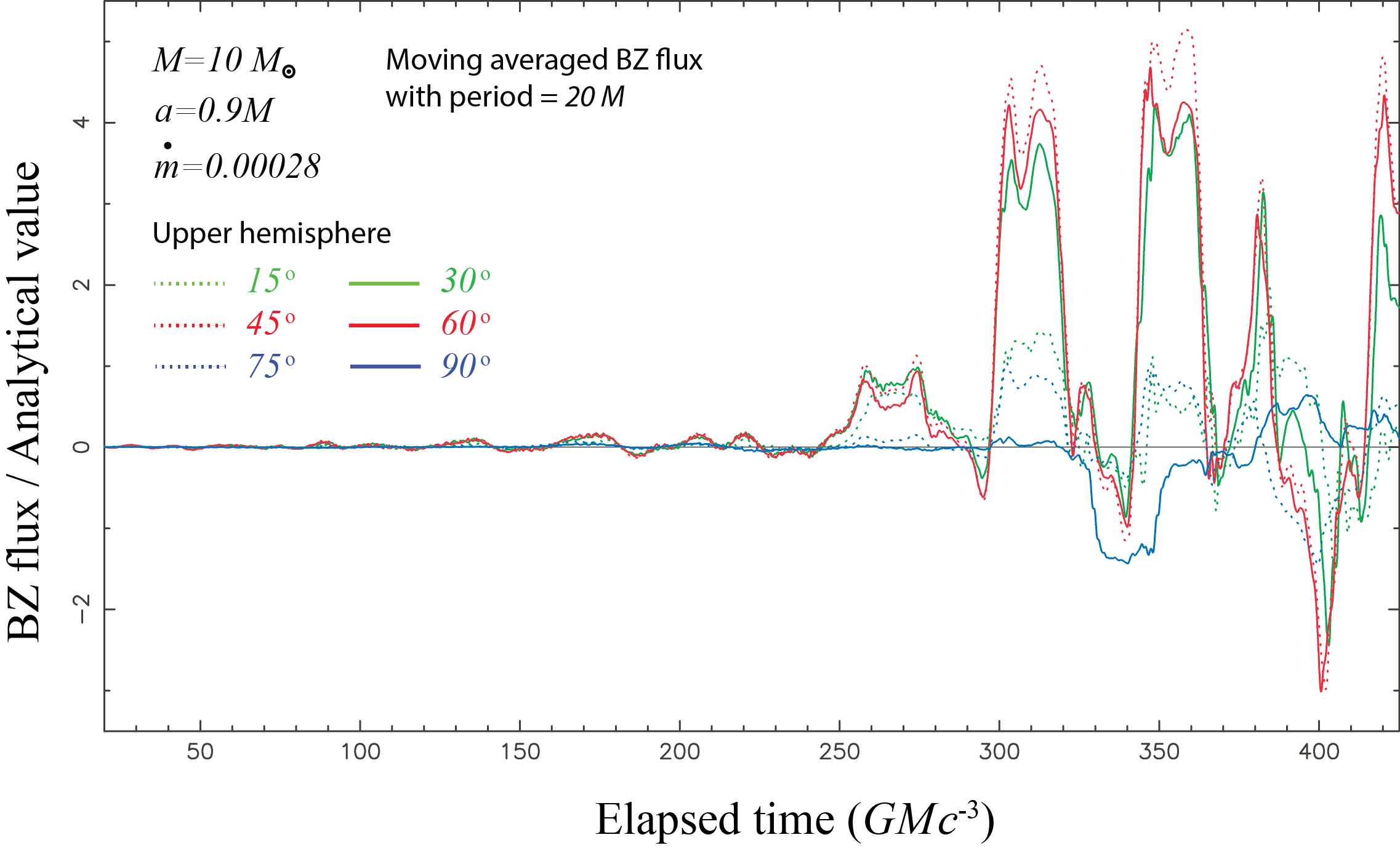}
\vspace*{-0.0truecm}
\caption{
% \textbf{
The Blandfor-Znajek flux moving-averaged with period $20M$
when $\dot{m}=2.8 \times 10^{-4}$, instead of $2.5 \times 10^{-4}$.
The flux is normalized by its analytical estimate (\S~\ref{sec:BZ_flux}),
and is measured at radius $r=2.335M$ and at six colatitudes
in the upper hemisphere as labelled.
% }
}
    \label{fig:LBZ_Time_MA_28}
\end{figure*}

%
% \textbf{
Let us briefly consider a greater accretion rate, 
$\dot{m} = 2.8 \times 10^{-4}$.
In figure~\ref{fig:LBZ_Time_MA_28},
we present moving-averaged BZ fluxes at six discrete colatitudes
for this case.
Comparing with the left panel of figure~\ref{fig:LBZ_Time_MA},
we find that the BZ flux shows flaring activity
with an interval $P_{\rm QPO} \approx 49M$,
and that the BZ flux concentrates in the middle latitudes
between $30^\circ$ and $60^\circ$.
Although the QPO frequency, $P_{\rm QPO}$, little depends on $\dot{m}$,
the peak of the normalized BZ flux decrease below $6$,
whereas it was between $8$ and $11$ when $\dot{m}=2.5 \times 10^{-4}$.
This is because the amplitude of the fluctuation reduces
owing to the increased plasma density
when $\dot{m}$ (and hence the pair production rate) increases.
% }

\begin{figure*}
%\vspace*{-16.0truecm}
\includegraphics[width=\columnwidth]{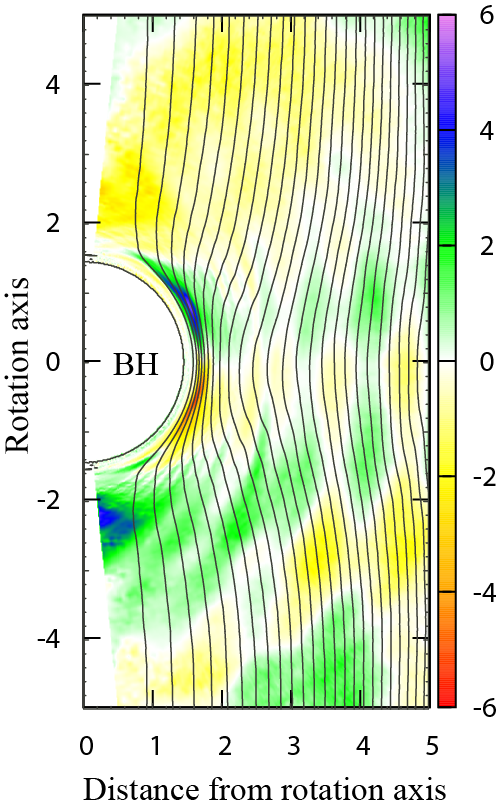}
\caption{
% \textbf{
Magnetic-field-aligned electric field,
$\mbox{\boldmath$E$} \cdot \mbox{\boldmath$B$} 
 / B_{\rm eq}(2M){}^2$ 
(color image)
on the poloidal plane ($r$,$\theta$),
when the BH's mass and spin parameter is 
$M=10M_\odot$ and $a=0.9M$, 
and the dimensionless mass accretion rate is
$\dot{m}=2.8 \times 10^{-4}$.
Black solid curves denote equi-$A_\varphi$ contours.
Both the abscissa and ordinate are measured in $GMc^{-2}$ unit.
The event horizon is located at $r=1.435M$.
% }
}
    \label{fig:Epara_3}
\end{figure*}

% \textbf{
As an example, we present 
$\mbox{\boldmath$E$} \cdot \mbox{\boldmath$B$} / B_{\rm eq}(2M){}^2$
at $t=300.00M$ (i.e., when the BZ flux peaks)
in figure~\ref{fig:Epara_3}.
Comparing with the top right panel of figure~\ref{fig:Epara_2},
which also shows $\mbox{\boldmath$E$} \cdot \mbox{\boldmath$B$}$
at a peak,
we find that the magnetic-field-aligned electric field
decreases with increasing $\dot{m}$.
In another word, the magnetosphere becomes highly charge-starved
when the pair production rate (eq.~[\ref{eq:dotN2}]) is small enough.
As $\dot{m}$ increases (i.e., as pair production increases), 
the magnetosphere approaches a force-free solution,
in the sense that the electric field is more efficiently
screened along the magnetic field lines.
However, if we adopt $\dot{m} \ge 3.0 \times 10^{-4}$,
the solutions become unstable from the horizon vicinity 
near the equator, 
because the variation of the compiled electromagnetic fields there 
becomes too sharp to be resolved even in the $r_\ast$ coordinates.
This is a limitation which we may encounter 
in the Boyer-Lindquist coordinates.
On the contrary, \citet{Parfrey:2019:PhRvL} 
adopted the Kerr-Schild coordinates,
which are regular at the horizon,
and demonstrated
that their PIC solutions approach the force-free solution
when they assumed greater pair-production rates than ours.
% }

\section{Discussion}
\label{sec:disc}
To sum up,
we simulated the evolution of a BH magnetosphere
by a PIC scheme,
when the electron-positron pair plasmas are steadily
supplied externally.
Provided that the mass accretion rate is as small as
$0.025$~\% of the Eddington limit,
the rotational energy of a black hole (BH)
is electromagnetically
extracted via the Blandford-Znajek process.
For a ten-solar-mass BH with the spin parameter $a=0.9M$,
the extracted energy flux
shows flaring activity with a period of 
$50$ dynamical time scales,
which is regulated by the magnetic reconnection
within the ergosphere.
During the flare, strong acceleration electric field
appears around the inner light surface
with a meridional return current towards the equator
inside the static limit.
The flare's energy flux concentrate along the
magnetic field lines that thread the event horizon
in the middle latitudes.
% However, when $\dot{m}$ decreases to $1 \times 10^{-4}$,
% the BZ flux becomes less oscillatory and exhibits
% weaker meridional concentration.

% Furthermore, we demonstrated that the BH magnetosphere
% becomes highly non-neutral within several Schwarzschild radii.
% Combining this result with 
% another analytical conclusion that 
% the Ohm's law completely breaks down in BH magnetospheres
% (Paper~I),
% we can conclude that the MHD approximation is not appropriate
% when we investigate highly vacuum magnetospheres 
% whose plasma density is comparable to the Goldreich-Julian value.
% Stellar-mass BHs in the low-hard or quiescent states,
% supermassive BHs in the center of low luminosity AGNs,
% and non-accreting pulsars possess such
% highly vacuum magnetospheres.
% For stellar-mass BHs,
% we demonstrated in \S~\ref{sec:ptcl}
% that the plasma density does take such low values
% when the mass accretion rate is much small compared to 
% the Eddington rate.

\subsection{Plasma skin depth}
\label{sec:skin}
Let us show that the plasma skin depth
(eq.~[\ref{eq:skin_depth}])
is resolved by the present grid interval.
Normalizing the pair density by the GJ value,
$n_\pm= \kappa n_{\rm GJ}$,
we obtain
\begin{eqnarray}
  \frac{l_{\rm p}}{r_{\rm g}}
  &=&
    \left( \frac{2 \langle\gamma\rangle}{\kappa} \right)^{1/2}
    \left( \frac{a}{M} \right)^{-1/2}
    \left( \frac{m}{m_{\rm e}} \right)^{1/2}
    \tilde{B}_0{}^{-1/2}
  \nonumber\\
  &=&
    1.44 
    \left( \frac{m}{m_{\rm p}/20} \right)^{1/2}
    \left( \frac{\gamma_7}{\kappa B_6 M_1}
           \frac{r_{\rm H}}{a}
    \right)^{1/2}
  \label{eq:skin_2}
\end{eqnarray}
where $\gamma_7 \equiv \langle\gamma\rangle / 10^7$
and $B_6 \equiv B / (10^6 \mbox{G})$;
$\tilde{B}_0 \equiv e B r_{\rm g} / m_{\rm e} c^2$
measures the magnetic field strength in dimensionless unit.
For a non-rotating BH (i.e., $a \rightarrow 0$), 
we find $\kappa \rightarrow \infty$
so that $\kappa n_{\rm GJ}$ gives the actual pair density.
In \citet{Parfrey:2019:PhRvL, Crinquand:2021:A&A, Bransgrove:2021:PhRvL},
they employed $\tilde{B}_0 < 5 \times 10^5$ in their re-scaled formulation,
which corresponds to $B < 5 \times 10^2$~G for $M=10 M_\odot$
and $B < 5 \times 10^{-5}$~G for $M=10^9 M_\odot$.
Note that we have
$B_{\rm eq} \sim 10^6 \dot{m}_{-4}{}^{1/2}$~G 
for $M=10 M_\odot$, and
$B_{\rm eq} \sim 10^2 \dot{m}_{-4}{}^{1/2}$~G 
for $M=10^9 M_\odot$,
where $\dot{m}_{-4} \equiv \dot{m}/10^{-4}$.
By virtue of such tiny magnetic field strengths,
$B \ll B_{\rm eq}$,
the skin depth was resolved in their works even for 
supermassive BHs, $M \sim 10^9 M_\odot$,
despite $\gamma_7 \ll 1$ due to the smallness of $B$.

In the present paper, on the other hand,
we adopted a stellar-size BH mass $M=10 M_\odot$ and
heavy electron/positron mass $m= m_{\rm p}/20$.
Evaluating the mean Lorentz factor, plasma density, 
and the magnetic field strength at each position at each time, 
we find $\gamma_7 / ( \kappa B_6 ) > 0.001$ at $r<3M$
and $\gamma_7 / ( \kappa B_6 ) > 0.006$ at $r>3M$
as the conservative lower limits, where $M= r_{\rm g}$.
Therefore, it follows from the last line of equation~(\ref{eq:skin_2})
that the plasma skin depth is greater than $0.045 M$ at $r<3M$,
which can be resolved by the present grid interval, $< 0.008 M$ there.
Also at $3M < r < 7M$,
we find that $l_{\rm p} > 0.11 M$ can be resolved 
by the grid interval $<0.026 M$ there.

\subsection{Horizontal magnetic field lines at the exact horizon}
\label{sec:horizontal_B}
As figure~\ref{fig:Blines} shows,
the magentic field lines become horizontal at the exact horizon
when magnetized plasmas accrete, 
if we adopt the Boyer-Lindquist coordinates.
Similar configuration of the lines of force also appear
for the electric field when two oppositely charged particles
are separated near the horizon
\citep[fig.~17 of \S~II~D~5 in][]{thorne86}.
Nevertheless, if we introduce the \lq stretched horizon'
slightly above the true horizon and consider the physics only
outside of the stretched horizon, 
we obtain horizon-penetrating lines of force at finite elapsed time.
After an infinite time elapses, 
the stretched horizon eventually matches the true horizon.

\begin{figure*}
%\hspace*{2.0truecm}
%\includegraphics[width=\columnwidth]{fig2.eps}
%\includegraphics[width=\textwidth]{fig3.pdf}
%\hspace{2.2cm}
\includegraphics[width=\textwidth, angle=0]{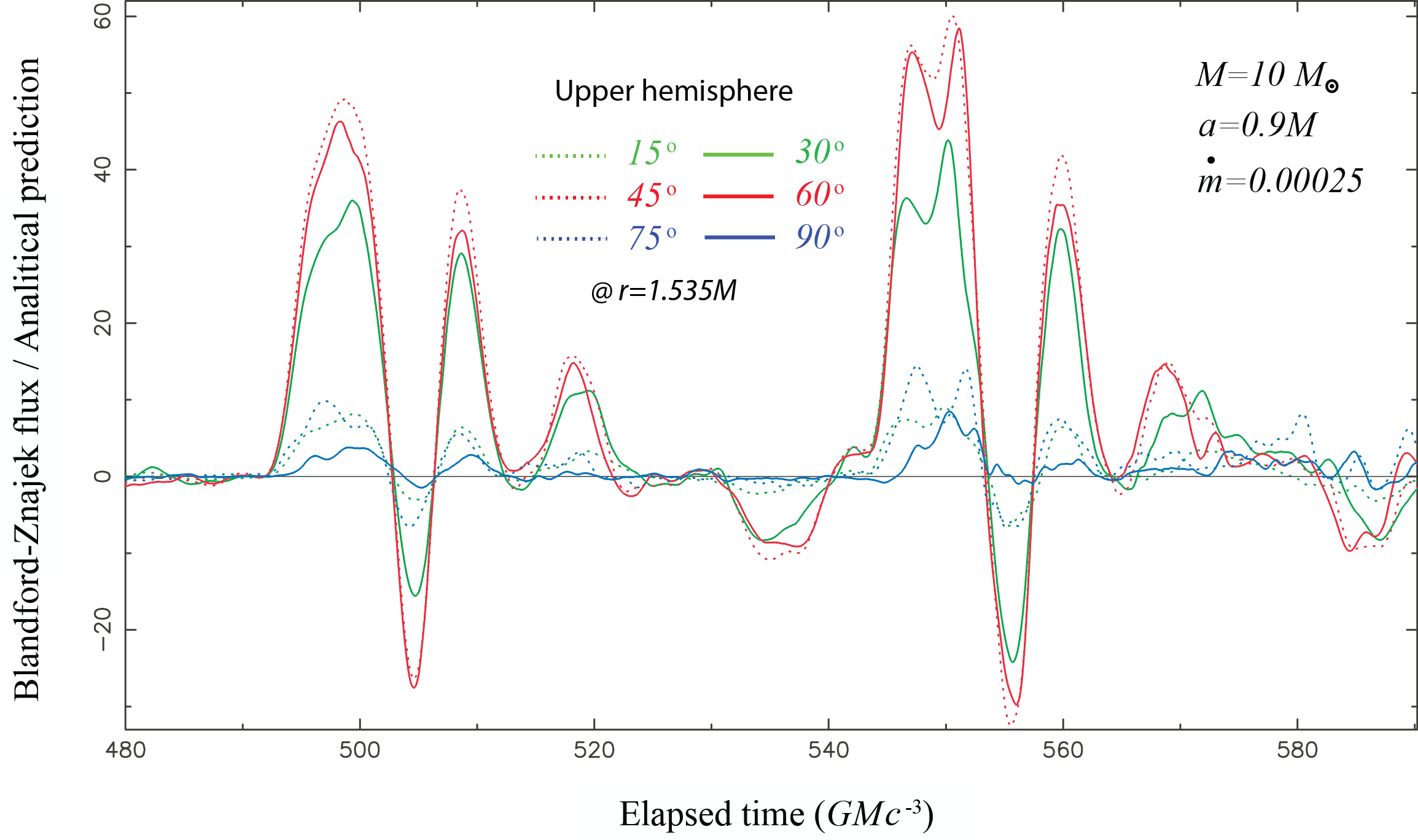}
\vspace*{-0.0truecm}
\caption{
Similar figure as the right panel of fig.~\ref{fig:LBZ_Time} 
(i.e., for $\dot{m}=2.5 \times 10^{-4}$)
but measured at a smaller radius, $r=1.535M$,
and at six discrete colatitudes as labeled.
}
    \label{fig:LBZ_Time_inner}
\end{figure*}

Although the time-dependent magnetic lines of force
is kept horizontal at the true horizon within a finite time,
it does not mean that the BH's rotational energy can be extracted
only at infinite elapsed time.
This is because the magnetized plasmas fall onto the horizon
with negative energies,
which realizes outward Poynting flux continuously in space.
In figure~\ref{fig:LBZ_Time_inner},
we present the BZ flux measured at much smaller radius 
$r= r_{\rm H} + 0.1M= 1.535M$.
Note that the coordinate time $t$ (i.e., the abscissa) corresponds to 
the proper time of a distant static observer.
By virtue of the time-dependent in-falling motion of 
magnetized plasmas with negative energies
\citep[see e.g.,][for the MHD Penrose proess]
{Hirotani:1992:ApJ,McKinney:2012:MNRAS},
comparable amount of energy is carried outward
at both $r=1.53M$ and $r=2.33M$
along individual magnetic flux tubes.

\subsection{Formation of limb-brightened jets}
\label{sec:limb}
Finally, let us briefly discuss how  
the middle-latitude concentration of the Poynting flux 
could result in the formation of
limb-brightened jets, which were observed from 
Mrk~501 \citep{Giroletti:2004:ApJ},
M87 \citep{hada16},
Cyg~A \citep{Boccardi:2016:A&A},
3C84 \citep{Kim:2019:A&A},
PG~1553+113 \citep{Lico:2020:A&A}, 
NGC315 \citep{Park:2021:ApJ}, and
Cen~A \citep{Janssen:2021:NatAs}.

In the present paper, 
we assumed that a nearly cylindrical
magnetic field lines are sustained by a strong, 
equatorial ring current at a large distance from the BH, 
and superposed the electromagnetic fields
created by the magnetospheric currents near the BH,
using a PIC method.
However, it is unrealistic to consider 
that such a ring current (that creates the Wald fields)
would exist at much larger distance 
than the jet downstream region.
Thus, it is natural that the external Wald field is
maintained only near the BH.
Nevertheless, we could concatenate
our PIC results (obtained near the BH)
with the downstream region (far away from the BH),
assuming e.g., that the time-averaged luminosity
(i.e., Poynting flux times cross section)
carried along each magnetic flux tube is constant
as a function of the distance from the BH.
Near the BH, we demonstrated that the Poynting flux is tiny
in the {\it higher} latitudes.
Thus, it is possible that a hollow jet is formed
in the sense that the Poynting flux is small 
along the jet axis.
In the subsequent paper,
we will convert the Poynting flux as a function of 
the magnetic flux function, $A_\varphi$, near the BH,
and quantitatively argue the formation of limb-brightened jets 
at large distances from the BH.

\bibliography{hirotani.bib}{}

\begin{thebibliography}{}
\expandafter\ifx\csname natexlab\endcsname\relax\def\natexlab#1{#1}\fi
\providecommand{\url}[1]{\href{#1}{#1}}
\providecommand{\dodoi}[1]{doi:~\href{http://doi.org/#1}{\nolinkurl{#1}}}
\providecommand{\doeprint}[1]{\href{http://ascl.net/#1}{\nolinkurl{http://ascl.net/#1}}}
\providecommand{\doarXiv}[1]{\href{https://arxiv.org/abs/#1}{\nolinkurl{https://arxiv.org/abs/#1}}}

\bibitem[{{Bardeen}(1970)}]{kerr63}
{Bardeen}, J.~M. 1970, \nat, 226, 64, \dodoi{10.1038/226064a0}

\bibitem[{{Beskin} {et~al.}(1992){Beskin}, {Istomin}, \& {Parev}}]{bes92}
{Beskin}, V.~S., {Istomin}, Y.~N., \& {Parev}, V.~I. 1992, \sovast, 36, 642

\bibitem[{{Blandford} \& {K{\"o}nigl}(1979)}]{Blandford:1979:ApJ}
{Blandford}, R.~D., \& {K{\"o}nigl}, A. 1979, \apj, 232, 34,
  \dodoi{10.1086/157262}

\bibitem[{{Blandford} \& {Znajek}(1977)}]{bla77}
{Blandford}, R.~D., \& {Znajek}, R.~L. 1977, \mnras, 179, 433,
  \dodoi{10.1093/mnras/179.3.433}

\bibitem[{{Boccardi} {et~al.}(2016){Boccardi}, {Krichbaum}, {Bach}, {Mertens},
  {Ros}, {Alef}, \& {Zensus}}]{Boccardi:2016:A&A}
{Boccardi}, B., {Krichbaum}, T.~P., {Bach}, U., {et~al.} 2016, \aap, 585, A33,
  \dodoi{10.1051/0004-6361/201526985}

\bibitem[{{Boyer} \& {Lindquist}(1967)}]{boyer67}
{Boyer}, R.~H., \& {Lindquist}, R.~W. 1967, Journal of Mathematical Physics, 8,
  265, \dodoi{10.1063/1.1705193}

\bibitem[{{Bransgrove} {et~al.}(2021){Bransgrove}, {Ripperda}, \&
  {Philippov}}]{Bransgrove:2021:PhRvL}
{Bransgrove}, A., {Ripperda}, B., \& {Philippov}, A. 2021, \prl, 127, 055101,
  \dodoi{10.1103/PhysRevLett.127.055101}

\bibitem[{{Chen} \& {Yuan}(2020)}]{Chen:2020:ApJ}
{Chen}, A.~Y., \& {Yuan}, Y. 2020, \apj, 895, 121,
  \dodoi{10.3847/1538-4357/ab8c46}

\bibitem[{{Crinquand} {et~al.}(2021){Crinquand}, {Cerutti}, {Dubus}, {Parfrey},
  \& {Philippov}}]{Crinquand:2021:A&A}
{Crinquand}, B., {Cerutti}, B., {Dubus}, G., {Parfrey}, K., \& {Philippov}, A.
  2021, \aap, 650, A163, \dodoi{10.1051/0004-6361/202040158}

\bibitem[{{Crinquand} {et~al.}(2020){Crinquand}, {Cerutti}, {Philippov},
  {Parfrey}, \& {Dubus}}]{Crinquand:2020:PhRvL}
{Crinquand}, B., {Cerutti}, B., {Philippov}, A.~e., {Parfrey}, K., \& {Dubus},
  G. 2020, \prl, 124, 145101, \dodoi{10.1103/PhysRevLett.124.145101}

\bibitem[{{Dhawan} {et~al.}(2000){Dhawan}, {Mirabel}, \&
  {Rodr{\'\i}guez}}]{Dhawan:2000:ApJ}
{Dhawan}, V., {Mirabel}, I.~F., \& {Rodr{\'\i}guez}, L.~F. 2000, \apj, 543,
  373, \dodoi{10.1086/317088}

\bibitem[{{Fender} {et~al.}(2004){Fender}, {Belloni}, \&
  {Gallo}}]{Fender:2004:MNRAS}
{Fender}, R.~P., {Belloni}, T.~M., \& {Gallo}, E. 2004, \mnras, 355, 1105,
  \dodoi{10.1111/j.1365-2966.2004.08384.x}

\bibitem[{{Ford} {et~al.}(2018){Ford}, {Keenan}, \&
  {Medvedev}}]{Ford:2018:PhRvD}
{Ford}, A.~L., {Keenan}, B.~D., \& {Medvedev}, M.~V. 2018, \prd, 98, 063016,
  \dodoi{10.1103/PhysRevD.98.063016}

\bibitem[{{Gallo} {et~al.}(2005){Gallo}, {Fender}, \&
  {Hynes}}]{Gallo:2005:MNRAS}
{Gallo}, E., {Fender}, R.~P., \& {Hynes}, R.~I. 2005, \mnras, 356, 1017,
  \dodoi{10.1111/j.1365-2966.2004.08503.x}

\bibitem[{{Giroletti} {et~al.}(2004){Giroletti}, {Giovannini}, {Feretti},
  {Cotton}, {Edwards}, {Lara}, {Marscher}, {Mattox}, {Piner}, \&
  {Venturi}}]{Giroletti:2004:ApJ}
{Giroletti}, M., {Giovannini}, G., {Feretti}, L., {et~al.} 2004, \apj, 600,
  127, \dodoi{10.1086/379663}

\bibitem[{{Hada} {et~al.}(2016){Hada}, {Kino}, {Doi}, {Nagai}, {Honma},
  {Akiyama}, {Tazaki}, {Lico}, {Giroletti}, {Giovannini}, {Orienti}, \&
  {Hagiwara}}]{hada16}
{Hada}, K., {Kino}, M., {Doi}, A., {et~al.} 2016, \apj, 817, 131,
  \dodoi{10.3847/0004-637X/817/2/131}

\bibitem[{{Hirotani}(2005)}]{Hirotani:2005:ApJ}
{Hirotani}, K. 2005, \apj, 619, 73, \dodoi{10.1086/426497}

\bibitem[{{Hirotani} {et~al.}(2021){Hirotani}, {Krasnopolsky}, {Shang},
  {Nishikawa}, \& {Watson}}]{Hirotani:2021:ApJ}
{Hirotani}, K., {Krasnopolsky}, R., {Shang}, H., {Nishikawa}, K.-i., \&
  {Watson}, M. 2021, \apj, 908, 88, \dodoi{10.3847/1538-4357/abd3a6}

\bibitem[{{Hirotani} \& {Okamoto}(1998)}]{Hirotani:1998:ApJ}
{Hirotani}, K., \& {Okamoto}, I. 1998, \apj, 497, 563, \dodoi{10.1086/305479}

\bibitem[{{Hirotani} \& {Pu}(2016)}]{hiro16a}
{Hirotani}, K., \& {Pu}, H.-Y. 2016, \apj, 818, 50,
  \dodoi{10.3847/0004-637X/818/1/50}

\bibitem[{{Hirotani} {et~al.}(2016){Hirotani}, {Pu}, {Lin}, {Chang}, {Inoue},
  {Kong}, {Matsushita}, \& {Tam}}]{Hirotani:2016:ApJ}
{Hirotani}, K., {Pu}, H.-Y., {Lin}, L. C.-C., {et~al.} 2016, \apj, 833, 142,
  \dodoi{10.3847/1538-4357/833/2/142}

\bibitem[{{Hirotani} {et~al.}(2017){Hirotani}, {Pu}, {Lin}, {Kong},
  {Matsushita}, {Asada}, {Chang}, \& {Tam}}]{Hirotani:2017:ApJ}
---. 2017, \apj, 845, 77, \dodoi{10.3847/1538-4357/aa7895}

\bibitem[{{Hirotani} {et~al.}(2018){Hirotani}, {Pu}, {Outmani}, {Huang}, {Kim},
  {Song}, {Matsushita}, \& {Kong}}]{Hirotani:2018:ApJ}
{Hirotani}, K., {Pu}, H.-Y., {Outmani}, S., {et~al.} 2018, \apj, 867, 120,
  \dodoi{10.3847/1538-4357/aae47a}

\bibitem[{{Hirotani} {et~al.}(1992){Hirotani}, {Takahashi}, {Nitta}, \&
  {Tomimatsu}}]{Hirotani:1992:ApJ}
{Hirotani}, K., {Takahashi}, M., {Nitta}, S.-Y., \& {Tomimatsu}, A. 1992, \apj,
  386, 455, \dodoi{10.1086/171031}

\bibitem[{{Hjellming} \& {Johnston}(1988)}]{Hjellming:1988:ApJ}
{Hjellming}, R.~M., \& {Johnston}, K.~J. 1988, \apj, 328, 600,
  \dodoi{10.1086/166318}

\bibitem[{{Ichimaru}(1977)}]{ichimaru77}
{Ichimaru}, S. 1977, \apj, 214, 840, \dodoi{10.1086/155314}

\bibitem[{{Jackson}(1962)}]{jackson62}
{Jackson}, J.~D. 1962, {Classical Electrodynamics}

\bibitem[{{Janssen} {et~al.}(2021){Janssen}, {Falcke}, {Kadler}, {Ros},
  {Wielgus}, {Akiyama}, {Balokovi{\'c}}, {Blackburn}, {Bouman}, {Chael},
  {Chan}, {Chatterjee}, {Davelaar}, {Edwards}, {Fromm}, {G{\'o}mez}, {Goddi},
  {Issaoun}, {Johnson}, {Kim}, {Koay}, {Krichbaum}, {Liu}, {Liuzzo}, {Markoff},
  {Markowitz}, {Marrone}, {Mizuno}, {M{\"u}ller}, {Ni}, {Pesce},
  {Ramakrishnan}, {Roelofs}, {Rygl}, {van Bemmel}, {Event Horizon Telescope
  Collaboration}, {Alberdi}, {Alef}, {Algaba}, {Anantua}, {Asada}, {Azulay},
  {Baczko}, {Ball}, {Ball}, {Barrett}, {Benson}, {Bintley}, {Bintley},
  {Blundell}, {Boland}, {Boland}, {Bower}, {Boyce}, {Bremer}, {Brinkerink},
  {Brissenden}, {Britzen}, {Broderick}, {Broguiere}, {Bronzwaer}, {Byun},
  {Carlstrom}, {Carlstrom}, {Carlstrom}, {Carlstrom}, {Chatterjee}, {Chen},
  {Chen}, {Chesler}, {Cho}, {Christian}, {Conway}, {Cordes}, {Crawford},
  {Crew}, {Cruz-Osorio}, {Cui}, {Cui}, {De Laurentis}, {Deane}, {Dempsey},
  {Desvignes}, {Dexter}, {Doeleman}, {Eatough}, {Farah}, {Farah}, {Fish},
  {Fomalont}, {Ford}, {Fraga-Encinas}, {Friberg}, {Friberg}, {Fuentes},
  {Galison}, {Gammie}, {Garc{\'\i}a}, {Gelles}, {Gentaz}, {Georgiev},
  {Georgiev}, {Gold}, {Gold}, {G{\'o}mez-Ruiz}, {Gu}, {Gurwell}, {Hada},
  {Haggard}, {Hecht}, {Hesper}, {Himwich}, {Ho}, {Ho}, {Honma}, {Huang},
  {Huang}, {Hughes}, {Ikeda}, {Inoue}, {James}, {Jannuzi}, {Jeter}, {Jiang},
  {Jimenez-Rosales}, {Jorstad}, {Jung}, {Karami}, {Karuppusamy}, {Kawashima},
  {Keating}, {Kettenis}, {Kim}, {Kim}, {Kim}, {Kino}, {Kino}, {Kofuji},
  {Koyama}, {Kramer}, {Kramer}, {Kuo}, {Lauer}, {Lee}, {Levis}, {Li}, {Li},
  {Lindqvist}, {Lico}, {Lindahl}, {Lindahl}, {Liu}, {Lo}, {Lobanov}, {Loinard},
  {Lonsdale}, {Lu}, {MacDonald}, {Mao}, {Marchili}, {Marscher},
  {Mart{\'\i}-Vidal}, {Matsushita}, {Matthews}, {Medeiros}, {Menten}, {Mizuno},
  {Moran}, {Moriyama}, {Moscibrodzka}, {Musoke}, {Mej{\'\i}as}, {Nagai},
  {Nagar}, {Nakamura}, {Narayan}, {Narayanan}, {Natarajan}, {Nathanail},
  {Neilsen}, {Neri}, {Noutsos}, {Nowak}, {Okino}, {Olivares}, {Ortiz-Le{\'o}n},
  {Oyama}, {{\"O}zel}, {Palumbo}, {Park}, {Patel}, {Pen}, {Pi{\'e}tu},
  {Plambeck}, {PopStefanija}, {Porth}, {P{\"o}tzl}, {Prather},
  {Preciado-L{\'o}pez}, {Psaltis}, {Pu}, {Rao}, {Rawlings}, {Raymond},
  {Rezzolla}, {Ricarte}, {Ripperda}, {Rogers}, {Rose}, {Roshanineshat},
  {Rottmann}, {Roy}, {Ruszczyk}, {S{\'a}nchez}, {S{\'a}nchez-Arguelles},
  {Sasada}, {Savolainen}, {Schloerb}, {Schuster}, {Shao}, {Shen}, {Small},
  {Sohn}, {SooHoo}, {Sun}, {Tazaki}, {Tetarenko}, {Tiede}, {Tilanus}, {Titus},
  {Torne}, {Trent}, {Traianou}, {Trippe}, {van Bemmel}, {van Langevelde}, {van
  Rossum}, {Wagner}, {Ward-Thompson}, {Wardle}, {Weintroub}, {Wex}, {Wharton},
  {Wharton}, {Wong}, {Wu}, {Yoon}, {Young}, {Young}, {Younsi}, {Yuan}, {Yuan},
  {Zensus}, {Zhao}, \& {Zhao}}]{Janssen:2021:NatAs}
{Janssen}, M., {Falcke}, H., {Kadler}, M., {et~al.} 2021, Nature Astronomy,
  \dodoi{10.1038/s41550-021-01417-w}

\bibitem[{{Jones} {et~al.}(1974{\natexlab{a}}){Jones}, {O'Dell}, \&
  {Stein}}]{Jones:1974a:ApJ}
{Jones}, T.~W., {O'Dell}, S.~L., \& {Stein}, W.~A. 1974{\natexlab{a}}, \apj,
  188, 353, \dodoi{10.1086/152724}

\bibitem[{{Jones} {et~al.}(1974{\natexlab{b}}){Jones}, {O'Dell}, \&
  {Stein}}]{Jones:1974b:ApJ}
---. 1974{\natexlab{b}}, \apj, 192, 261, \dodoi{10.1086/153057}

\bibitem[{{Katsoulakos} \& {Rieger}(2020)}]{Katsoulakos:2020:ApJ}
{Katsoulakos}, G., \& {Rieger}, F.~M. 2020, \apj, 895, 99,
  \dodoi{10.3847/1538-4357/ab8fa1}

\bibitem[{{Kim} {et~al.}(2019){Kim}, {Krichbaum}, {Marscher}, {Jorstad},
  {Agudo}, {Thum}, {Hodgson}, {MacDonald}, {Ros}, {Lu}, {Bremer}, {de Vicente},
  {Lindqvist}, {Trippe}, \& {Zensus}}]{Kim:2019:A&A}
{Kim}, J.~Y., {Krichbaum}, T.~P., {Marscher}, A.~P., {et~al.} 2019, \aap, 622,
  A196, \dodoi{10.1051/0004-6361/201832920}

\bibitem[{{Kino} {et~al.}(2014){Kino}, {Takahara}, {Hada}, \&
  {Doi}}]{Kino:2014:ApJ}
{Kino}, M., {Takahara}, F., {Hada}, K., \& {Doi}, A. 2014, \apj, 786, 5,
  \dodoi{10.1088/0004-637X/786/1/5}

\bibitem[{{Kisaka} {et~al.}(2020){Kisaka}, {Levinson}, \&
  {Toma}}]{Kisaka:2020:ApJ}
{Kisaka}, S., {Levinson}, A., \& {Toma}, K. 2020, \apj, 902, 80,
  \dodoi{10.3847/1538-4357/abb46c}

\bibitem[{{Koide} {et~al.}(2002){Koide}, {Shibata}, {Kudoh}, \&
  {Meier}}]{Koide:2002:Sci}
{Koide}, S., {Shibata}, K., {Kudoh}, T., \& {Meier}, D.~L. 2002, Science, 295,
  1688, \dodoi{10.1126/science.1068240}

\bibitem[{{Langdon} \& {Lasinski}(1976)}]{Langdon:1976:cofu}
{Langdon}, A.~B., \& {Lasinski}, B.~F. 1976, in Controlled Fusion, 327--366

\bibitem[{{Levinson}(2000)}]{Levinson:2000:PhRvL}
{Levinson}, A. 2000, \prl, 85, 912, \dodoi{10.1103/PhysRevLett.85.912}

\bibitem[{{Levinson} \& {Cerutti}(2018)}]{Levinson:2018:AA}
{Levinson}, A., \& {Cerutti}, B. 2018, \aap, 616, A184,
  \dodoi{10.1051/0004-6361/201832915}

\bibitem[{{Levinson} \& {Rieger}(2011)}]{levi11}
{Levinson}, A., \& {Rieger}, F. 2011, \apj, 730, 123,
  \dodoi{10.1088/0004-637X/730/2/123}

\bibitem[{{Levinson} \& {Segev}(2017)}]{Levinson:2017:PhRvD}
{Levinson}, A., \& {Segev}, N. 2017, \prd, 96, 123006,
  \dodoi{10.1103/PhysRevD.96.123006}

\bibitem[{{Lico} {et~al.}(2020){Lico}, {Liu}, {Giroletti}, {Orienti},
  {G{\'o}mez}, {Piner}, {MacDonald}, {D'Ammando}, \& {Fuentes}}]{Lico:2020:A&A}
{Lico}, R., {Liu}, J., {Giroletti}, M., {et~al.} 2020, \aap, 634, A87,
  \dodoi{10.1051/0004-6361/201936564}

\bibitem[{{Lobanov}(1998)}]{Lobanov:1998:A&A}
{Lobanov}, A.~P. 1998, \aap, 330, 79.
\newblock \doarXiv{astro-ph/9712132}

\bibitem[{{Mahadevan}(1997)}]{Mahadevan:1997:ApJ}
{Mahadevan}, R. 1997, \apj, 477, 585, \dodoi{10.1086/303727}

\bibitem[{{Marscher}(1983)}]{Marscher:1983:ApJ}
{Marscher}, A.~P. 1983, \apj, 264, 296, \dodoi{10.1086/160597}

\bibitem[{{McKinney} {et~al.}(2012){McKinney}, {Tchekhovskoy}, \& {Bland
  ford}}]{McKinney:2012:MNRAS}
{McKinney}, J.~C., {Tchekhovskoy}, A., \& {Bland ford}, R.~D. 2012, \mnras,
  423, 3083, \dodoi{10.1111/j.1365-2966.2012.21074.x}

\bibitem[{{Mo{\'s}cibrodzka} {et~al.}(2011){Mo{\'s}cibrodzka}, {Gammie},
  {Dolence}, \& {Shiokawa}}]{Moscibrodzka:2011:ApJ}
{Mo{\'s}cibrodzka}, M., {Gammie}, C.~F., {Dolence}, J.~C., \& {Shiokawa}, H.
  2011, \apj, 735, 9, \dodoi{10.1088/0004-637X/735/1/9}

\bibitem[{{Narayan} {et~al.}(2021){Narayan}, {Chael}, {Chatterjee}, {Ricarte},
  \& {Curd}}]{Narayan:2021:arXiv}
{Narayan}, R., {Chael}, A., {Chatterjee}, K., {Ricarte}, A., \& {Curd}, B.
  2021, arXiv e-prints, arXiv:2108.12380.
\newblock \doarXiv{2108.12380}

\bibitem[{{Narayan} \& {Yi}(1994)}]{narayan94}
{Narayan}, R., \& {Yi}, I. 1994, \apjl, 428, L13, \dodoi{10.1086/187381}

\bibitem[{{Nishikawa} {et~al.}(2021){Nishikawa}, {Du{\r{A}}{\textsterling}an},
  {K{\"o}hn}, \& {Mizuno}}]{Nishikawa:2021:LRCA}
{Nishikawa}, K., {Du{\r{A}}{\textsterling}an}, I., {K{\"o}hn}, C., \& {Mizuno},
  Y. 2021, Living Reviews in Computational Astrophysics, 7, 1,
  \dodoi{10.1007/s41115-021-00012-0}

\bibitem[{{Parfrey} {et~al.}(2019){Parfrey}, {Philippov}, \&
  {Cerutti}}]{Parfrey:2019:PhRvL}
{Parfrey}, K., {Philippov}, A., \& {Cerutti}, B. 2019, \prl, 122, 035101,
  \dodoi{10.1103/PhysRevLett.122.035101}

\bibitem[{{Park} {et~al.}(2021){Park}, {Hada}, {Nakamura}, {Asada}, {Zhao}, \&
  {Kino}}]{Park:2021:ApJ}
{Park}, J., {Hada}, K., {Nakamura}, M., {et~al.} 2021, \apj, 909, 76,
  \dodoi{10.3847/1538-4357/abd6ee}

\bibitem[{{Ptitsyna} \& {Neronov}(2016)}]{Ptitsyna:2016:A&A}
{Ptitsyna}, K., \& {Neronov}, A. 2016, \aap, 593, A8,
  \dodoi{10.1051/0004-6361/201527549}

\bibitem[{{Punsly}(1996)}]{Punsly:1996:ApJ}
{Punsly}, B. 1996, \apj, 467, 105, \dodoi{10.1086/177588}

\bibitem[{{Reynolds} {et~al.}(1996){Reynolds}, {Fabian}, {Celotti}, \&
  {Rees}}]{Reynolds:1996:MNRAS}
{Reynolds}, C.~S., {Fabian}, A.~C., {Celotti}, A., \& {Rees}, M.~J. 1996,
  \mnras, 283, 873, \dodoi{10.1093/mnras/283.3.873}

\bibitem[{{Stirling} {et~al.}(2001){Stirling}, {Spencer}, {de la Force},
  {Garrett}, {Fender}, \& {Ogley}}]{Stirling:2001:MNRAS}
{Stirling}, A.~M., {Spencer}, R.~E., {de la Force}, C.~J., {et~al.} 2001,
  \mnras, 327, 1273, \dodoi{10.1046/j.1365-8711.2001.04821.x}

\bibitem[{{Tchekhovskoy} {et~al.}(2010){Tchekhovskoy}, {Narayan}, \&
  {McKinney}}]{Tchekhovskoy:2010:ApJ}
{Tchekhovskoy}, A., {Narayan}, R., \& {McKinney}, J.~C. 2010, \apj, 711, 50,
  \dodoi{10.1088/0004-637X/711/1/50}

\bibitem[{{Tchekhovskoy} {et~al.}(2011){Tchekhovskoy}, {Narayan}, \&
  {McKinney}}]{Tchekhovskoy:2011:MNRAS}
---. 2011, \mnras, 418, L79, \dodoi{10.1111/j.1745-3933.2011.01147.x}

\bibitem[{{Thorne} {et~al.}(1986){Thorne}, {Price}, \& {MacDonald}}]{thorne86}
{Thorne}, K.~S., {Price}, R.~H., \& {MacDonald}, D.~A. 1986, {Black holes: The
  membrane paradigm}

\bibitem[{{Villasenor} \& {Buneman}(1992)}]{villa92}
{Villasenor}, J., \& {Buneman}, O. 1992, Computer Physics Communications, 69,
  306, \dodoi{10.1016/0010-4655(92)90169-Y}

\bibitem[{{Wald}(1974)}]{Wald:1974:PhRvD}
{Wald}, R.~M. 1974, \prd, 10, 1680, \dodoi{10.1103/PhysRevD.10.1680}

\bibitem[{{Wardle} {et~al.}(1998){Wardle}, {Homan}, {Ojha}, \&
  {Roberts}}]{Wardle:1998:Natur}
{Wardle}, J.~F.~C., {Homan}, D.~C., {Ojha}, R., \& {Roberts}, D.~H. 1998, \nat,
  395, 457, \dodoi{10.1038/26675}

\bibitem[{{Wong} {et~al.}(2021){Wong}, {Ryan}, \& {Gammie}}]{Wong:2021:ApJ}
{Wong}, G.~N., {Ryan}, B.~R., \& {Gammie}, C.~F. 2021, \apj, 907, 73,
  \dodoi{10.3847/1538-4357/abd0f9}

\bibitem[{{Yuan} \& {Narayan}(2014)}]{Yuan:2014:ARA&A}
{Yuan}, F., \& {Narayan}, R. 2014, \araa, 52, 529,
  \dodoi{10.1146/annurev-astro-082812-141003}

\end{thebibliography}
\bibliographystyle{aasjournal}



\end{document}